\documentclass[sigconf]{acmart}

\AtBeginDocument{%
  \providecommand\BibTeX{{%
    \normalfont B\kern-0.5em{\scshape i\kern-0.25em b}\kern-0.8em\TeX}}}

\newcommand{\QUOTE}[1]{\textsf{\textit{\textcolor{black}{``#1''}}}}
\newcommand{\CHANGE}[1]{\textcolor{black}{#1}}

\newcommand{\PROMPT}[1]{%
    \begingroup
    \spaceskip=0.2em plus 0.1em minus 0.1em 
    \textsf{\textcolor{black}{\small{\fontfamily{cmss}\selectfont ``#1''}}}%
    \endgroup
}

\copyrightyear{2024}
\acmYear{2024}
\setcopyright{acmlicensed}\acmConference[ASSETS '24]{The 26th International ACM SIGACCESS Conference on Computers and Accessibility}{October 27--30, 2024}{St. John's, NL, Canada}
\acmBooktitle{The 26th International ACM SIGACCESS Conference on Computers and Accessibility (ASSETS '24), October 27--30, 2024, St. John's, NL, Canada}
\acmDOI{10.1145/3663548.3675599}
\acmISBN{979-8-4007-0677-6/24/10}

\usepackage{hhline}
\usepackage{multirow}
\usepackage{makecell}
\usepackage{fancyvrb}
\usepackage{xcolor}
\usepackage{listings}
\usepackage{enumitem}

\colorlet{punct}{red!60!black}
\definecolor{background}{HTML}{EEEEEE}
\definecolor{delim}{RGB}{20,105,176}
\colorlet{numb}{magenta!60!black}

\lstdefinelanguage{json}{
    basicstyle=\normalfont\ttfamily,
    showstringspaces=false,
    breaklines=true,
    frame=lines,
    literate=
     *{0}{{{\color{numb}0}}}{1}
      {1}{{{\color{numb}1}}}{1}
      {2}{{{\color{numb}2}}}{1}
      {3}{{{\color{numb}3}}}{1}
      {4}{{{\color{numb}4}}}{1}
      {5}{{{\color{numb}5}}}{1}
      {6}{{{\color{numb}6}}}{1}
      {7}{{{\color{numb}7}}}{1}
      {8}{{{\color{numb}8}}}{1}
      {9}{{{\color{numb}9}}}{1}
      {:}{{{\color{punct}{:}}}}{1}
      {,}{{{\color{punct}{,}}}}{1}
      {\{}{{{\color{delim}{\{}}}}{1}
      {\}}{{{\color{delim}{\}}}}}{1}
      {[}{{{\color{delim}{[}}}}{1}
      {]}{{{\color{delim}{]}}}}{1},
}

\def\name{EditScribe}

\begin{document}
\author{Ruei-Che Chang}
\email{rueiche@umich.edu}
\affiliation{%
 \institution{University of Michigan}
 \city{Ann Arbor, MI}
 \country{USA}
}

\author{Yuxuan Liu}
\email{liurick@umich.edu}
\affiliation{%
 \institution{University of Michigan}
 \city{Ann Arbor, MI}
 \country{USA}
}

\author{Lotus Zhang}
\email{hanziz@uw.edu}
\affiliation{%
 \institution{University of Washington}
 \city{Seattle, Washington}
 \country{USA}
}

\author{Anhong Guo}
\email{anhong@umich.edu}
\affiliation{%
 \institution{University of Michigan}
 \city{Ann Arbor, MI}
 \country{USA}
}

\title[{\name}]{ {\name}: Non-Visual Image Editing with \\ Natural Language Verification Loops}

\renewcommand{\shortauthors}{Ruei-Che Chang, Yuxuan Liu, Lotus Zhang and Anhong Guo}

\begin{abstract}
Image editing is an iterative process that requires precise visual evaluation and manipulation for the output to match the editing intent. 
However, current image editing tools do not provide accessible interaction nor sufficient feedback for blind and low vision individuals to achieve this level of control.
To address this, we developed \textit{{\name}}, a prototype system that makes object-level image editing actions accessible using \textit{natural language verification loops} powered by large multimodal models.
Using {\name}, the user first comprehends the image content through initial general and object descriptions, then specifies edit actions using open-ended natural language prompts. {\name} performs the image edit, and provides four types of verification feedback for the user to verify the performed edit, including a summary of visual changes, AI judgement, and updated general and object descriptions. The user can ask follow-up questions to clarify and probe into the edits or verification feedback, before performing another edit.
In a study with ten blind or low-vision users, we found that {\name} supported participants to perform and verify image edit actions non-visually. We observed different prompting strategies from participants, and their perceptions on the various types of verification feedback.
Finally, we discuss the implications of leveraging natural language verification loops to make visual authoring non-visually accessible.

\end{abstract}



\begin{teaserfigure}
  \includegraphics[width=\textwidth]{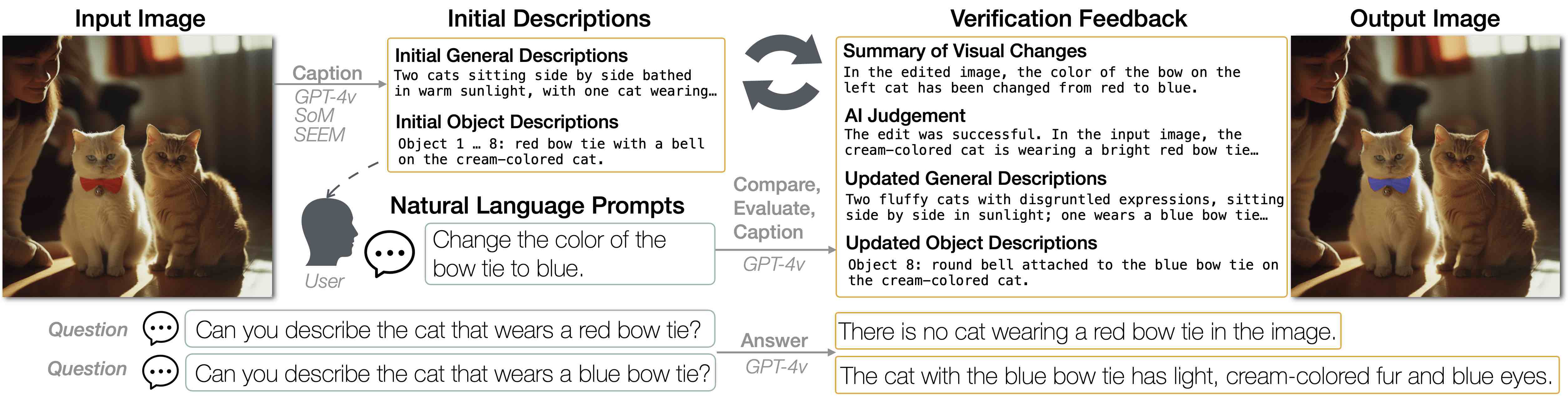}
  \caption{{
  \name} supports non-visual image editing using natural language verification loops.
  The user first comprehends the image content through initial general and object descriptions, then specifies edit actions using natural language. 
  {\name} performs the image edit, and provides four types of verification feedback for the user to verify the performed edit, including a \textit{summary of visual changes}, \textit{AI judgement}, and \textit{updated general and object descriptions}.
  The user can ask follow-up questions to clarify and probe into the edits or verification feedback, before performing another edit.
  }
  \Description{An input image showing the two cats, one is white with red bow tie and another orange one, and a woman kneeling on the floor. This image is pointed to initial description through the arrow.
The user first comprehends the image content through initial general and object descriptions, generated by GPT-4v
(CONTENT: Initial general descriptions: Two cats sitting side by side bathed in warm sunlight, with one cat wearing…, Initial Object descriptions: Object 1…8: red bow tie with a bell on the cream-colored cat).
The user then specifies editing actions using natural language (QUERY: Change the color of the bow tie to blue). 
The user obtain four types of verification feedback such as summary of visual changes (CONTENT: In the edited image, the color of the bow on the left cat  has been changed from red to blue), AI judgement (CONTENT: The edit was successful. In the input image, the cream-colored cat is wearing a bright red bow tie…), Updated general descriptions (CONTENT: Two lfuffly cats with disgruntled expressions, sitting side by side in sunlight; one wears a blue bow tie…), and updated object descriptions (CONTENT: Object 8: round bell attached to the blue bow tie on the cream-colored cat.)
The user can ask follow-up questions to clarify and probe into the edits or verification feedback, before performing another edit, such as “Can you describe the cat that wears red bow tie?” With GPT-4v answering “There is no cat wearing a red bow tie in the image”, and another question “Can you describe the cat that wears the blue bow tie?” With GPT-4v answering The cat with the blue bow tie has light, cream-colored fur and blue eyes.
An output image showing the same as input image but the blue bow tie was changed from red to blue.
}
  \label{fig:teaser}
\end{teaserfigure}

\maketitle

\section{Introduction}
Images are crucial visual media used to convey information and serve a variety of critical purposes in our everyday lives (\textit{e.g., }work, social, entertainment). 
Blind and low-vision (BLV) individuals often need to create, edit, and share images with sighted peers, for social engagement, visual information access, and many other personal pursuits~\cite{lotus23assets, sadji2021, dustin2013, schaadhardt2021, jun2021}. 
Crucial to producing a satisfactory image is reviewing and editing its visual details. 
For BLV individuals, the necessity for image editing arises in various contexts, such as enhancing photo quality (\textit{e.g., }clarity, lighting, and composition) in blind photography~\cite{bennett2018, gonzalez2022} and obfuscating private content \cite{Ahmed2016, alharbi2022cscw, PenuelaAssets22, zhang2023imageally, lotusChi24}. However, the BLV community commonly faces access barriers with the image editing process, as existing editing tools do not support comprehensive non-visual understanding of the image content nor accessible visual manipulation control for them to iteratively perform edits and evaluate the visual changes~\cite{voykinska2016, gonzalez2022, harada2013, dustin2013, lotusChi24}. 

Recently, HCI and accessibility researchers have increasingly focused on developing accessible tools for the BLV community in visual content creation and editing, including enabling the perception of elemental changes on presentation slide editing platforms \cite{diffscriber,collaba11y,a11yboard}, making video editing accessible through alternative textual representations for sounds and visuals \cite{avscript}, modifying layouts with tactile supports~\cite{li2019}, and allowing users to verify content created by generative AI models with preset verification prompts \cite{genassist}. 
Despite these efforts, the domain of image editing remains under-explored, 
which is visually challenging as it requires users to precisely understand image content, perform desired edit effects, and evaluate changes \cite{alharbi2022cscw,lotusChi24}. Thus in this work, we aim to address the following questions:
\begin{quote}\leftskip-18pt\rightskip-18pt
\textit{How to enable BLV people to perform image editing non-visually? And how to support the evaluation of visual changes after edits?}
\end{quote}

To achieve this, we propose \textit{{\name}}, a prototype system that utilizes large multimodal models (LMMs) to make object-level image editing actions non-visually accessible to BLV people. 
The core of {\name} is \textit{natural language verification loops}. 
Using {\name}, the user first comprehends the image content through initial general and object descriptions, then specifies edit actions using open-ended natural language prompts. 
{\name} performs the image edit, and provides four types of verification feedback for the user to verify the performed edit, including \textsc{Summary of Visual Changes}, \textsc{AI Judgement}, and updated \textsc{General} and \textsc{Object Descriptions}. 
The user can ask follow-up questions to clarify and probe into the edits or verification feedback before performing another edit.
This set of verification feedback was grounded on both visual and textual sources, such as comparing the images before and after an edit to summarize visual changes, using the edited image to provide new and independent image and object descriptions, and combining the above for AI to judge and reason about the success of the edit.
We demonstrated the utility of natural language verification loops with five object-level edit actions in {\name}, including blurring an object, removing an object, changing the color of an object, adjusting the brightness of the object, and adding text to the image.
We focus on object-level actions as they require precise understanding and manipulation of image details, and are critical to tasks commonly desired by BLV people \cite{alharbi2022cscw, bennett2018, voykinska2016}, \textit{e.g.,} image obfuscation for privacy \cite{alharbi2022cscw, lotusChi24} and background removal/blurring to focus on specific objects \cite{bennett2018, visphoto}.

We evaluated {\name} with 10 BLV participants to understand \textit{RQ1: How does {\name} support
non-visual image editing?} \textit{RQ2: How do BLV people prompt {\name}?} \textit{RQ3: How do BLV people perceive {\name}'s verification feedback?} and 
\textit{RQ4: How do BLV people perceive the final edited images?}
We found that BLV participants were able to perform most of the editing tasks using {\name}, and they had different strategies for prompting, such as creating prompts ranging from detailed to succinct to adapt to discrepant verification feedback, using varying tones or wordings to facilitate the system's understanding, or asking follow-up questions to gauge the reliability and build their trust in {\name}.
Also, participants had different preferences for each verification feedback, depending on the contexts of use. 
Other factors, such as inconsistent or discrepant information, visual experiences and expectations, and tones of verification feedback, also affected their trust and confidence in {\name}. 
Furthermore, participants felt confident about the edited images by {\name} and were willing to publish based on the context, but preferred a second check on the final edited images using sighted assistance or other AI services. 
Finally, we discussed how {\name} could be extended to support additional edit actions, provide enhanced verification feedback, 
and implications for future systems to leverage natural language verification loops for content creation.

\vspace{-0.5pc}
\section{Related work}
Our research is informed by prior work in visual content authoring accessibility, image accessibility, and AI-assisted image editing.
\vspace{-0.5pc}

\subsection{Visual Content Authoring Accessibility} \label{authoring_accessibility}
Research around BLV individuals’ digital creative activities has recently gained momentum in the field of accessibility. This body of research noted substantial interest from the community in a wide range of digital content creation~\cite{lotus23assets, dustin2013, saha2020, schaadhardt2021, jun2021}. Notably, BLV individuals commonly feel the need to engage in visual content authoring, including but not limited to photography, videos, presentation slides, data visualization, websites, and visual-heavy documents~\cite{lotus23assets, dustin2013, schaadhardt2021, jun2021}. These visual creative activities bring BLV individuals opportunities for information access, socializing, employment, self-expression, entertainment and more~\cite{lotus23assets}. For example, BLV people often share photos with sighted peers or remote agents (\textit{e.g., }Aira~\cite{aira}, Be My Eyes~\cite{bemyeyes}) for visual interpretation assistance~\cite{sadji2021}, while many also desire access to authoring visual content to better engage with popular social media~\cite{rong2022, bennett2018, lotus23assets}. 

However, in authoring visual content, BLV individuals face immense access barriers~\cite{bennett2018, gonzalez2022, park2020, harada2013, dustin2013}. Past surveys revealed that digital creative activities that involve visual elements are generally considered more challenging to BLV creators~\cite{lotus23assets}. 
While basic photography and document formatting tasks have been increasingly supported by technology~\cite{textanalyzer, bemyai, voiceover}, \textit{editing} visual elements remains particularly challenging~\cite{lotus23assets, park2020}---most often, BLV individuals shy away from visual editing tasks or rely on sighted support. 

Specific visual editing challenges to BLV individuals were also noted, including limited understanding of visual editing standards, insufficient information about the visuals, difficulties perceiving the effect of edits, and inaccessible editing controls~\cite{schaadhardt2021, potluri2022, potluri2021, li2019, lotus23assets}. 
Still, additional challenges arise across different editing contexts. For example, when editing out private or sensitive information, BLV individuals experience heightened uncertainty about how obscured the result is~\cite{alharbi2022cscw,lotusChi24}. 
In authoring content in collaboration, there are challenges with reviewing others' visual changes~\cite{diffscriber}. For editing visuals in motion---videos, blind creators also have difficulties skimming through visual information~\cite{huh2023}. 

Given how BLV individuals’ needs for visual editing support could vary greatly based on their personal experience and authoring contexts, we consider natural language interaction a potentially more accessible image editing method, 
as it could allow on-demand visual manipulation and iterative information-seeking. 
So far, natural language interaction has not been utilized for accessible image editing, and we know little about how to design such interaction. 
To address this research gap, we developed and evaluated \textit{{\name}}, a system that utilizes LMMs for non-visual, text-to-image editing make image editing. 
We further explored how BLV users make use of natural language interaction in different contexts of image editing as well as their design feedback.

\subsection{BLV Individuals' Access to Images}
Key to successful image editing is the access to relevant visual information in the image, which most BLV screen reader users perceive through \CHANGE{image descriptions}~\cite{w3cimageconcepts, alttxt, stangl21assets}. A high-quality image description should provide a comparable experience to encountering the image visually~\cite{stangl21assets} (\textit{e.g., }providing the purpose of the image, object or people present, etc.~\cite{w3cimageconcepts, petrie2005describing, diagramcenter2015, w3coverview}).
Also, it is critical to create image descriptions that adapt to individuals' information-seeking goals \cite{wu2017, diagramcenter2015, w3coverview} but not ``one size fits all'' \cite{stangl21assets}.
For example, for visual elements on social media images, information related to specific persons, location, photo quality and others’ responses are all important to describe~\cite{wu2017, zhao2017, voykinska2016, morris2016}, while for data visualization, granular and objective descriptions (\textit{e.g., }shape, context) could support users’ interpretation freedom~\cite{fan2022}. 
To increase the availability of alternative texts online, AI is considered as a scalable solution~\cite{wu2017, zhao2017, macleod2017, bigham2010}. Automated image descriptions and object detection tools have become increasingly available (e.g.,~\cite{bemyai, seeingai}). 
However, as state-of-the-art AI models still produce inaccurate results, these systems need to provide enough information for users to notice potential undesired outcomes, such as by clearly communicating what the system is able to do and how well it could perform the intended task~\cite{wu2017, zhao2017, macleod2017, amershi2019}. 

In addition to the above guidelines, providing image description in the context of visual content \textit{sharing and editing} involves additional considerations. In authoring visual content, BLV individuals tend to desire more information about their images and description accuracy~\cite{jung2022, zhao2017, schaadhardt2021}. In particular, spatial layout, appearances of objects, and timely feedback to visual changes critically influence BLV creators’ judgement of necessary edits~\cite{schaadhardt2021, diffscriber, lotusChi24, genassist}. Image descriptions also need to include information specific to the editing needs in different content authoring contexts (as mentioned in Section~\ref{authoring_accessibility}), \textit{e.g., } focused descriptions on potentially private objects for managing the visual privacy~\cite{alharbi2022cscw,lotusChi24}, visual changes for collaborative authoring~\cite{diffscriber}. 

Image descriptions for content authoring purposes thus tend to involve an abundance of visual details, a key challenge that needs to be designed around~\cite{li2019, schaadhardt2021}. 
Still, investigation on image descriptions in the context of visual editing is limited, especially around supporting BLV creators' understanding of visual changes on images~\cite{lotusChi24}. 
In this paper, we draw on prior work and design a set of verification feedback on an image edit. 
We use it as a probe to gain more insights into how BLV users perceived it to inform potential improvements.

\begin{figure*}[t]
\begin{center}
\vspace{-1.pc}
\includegraphics[width=0.9\linewidth]{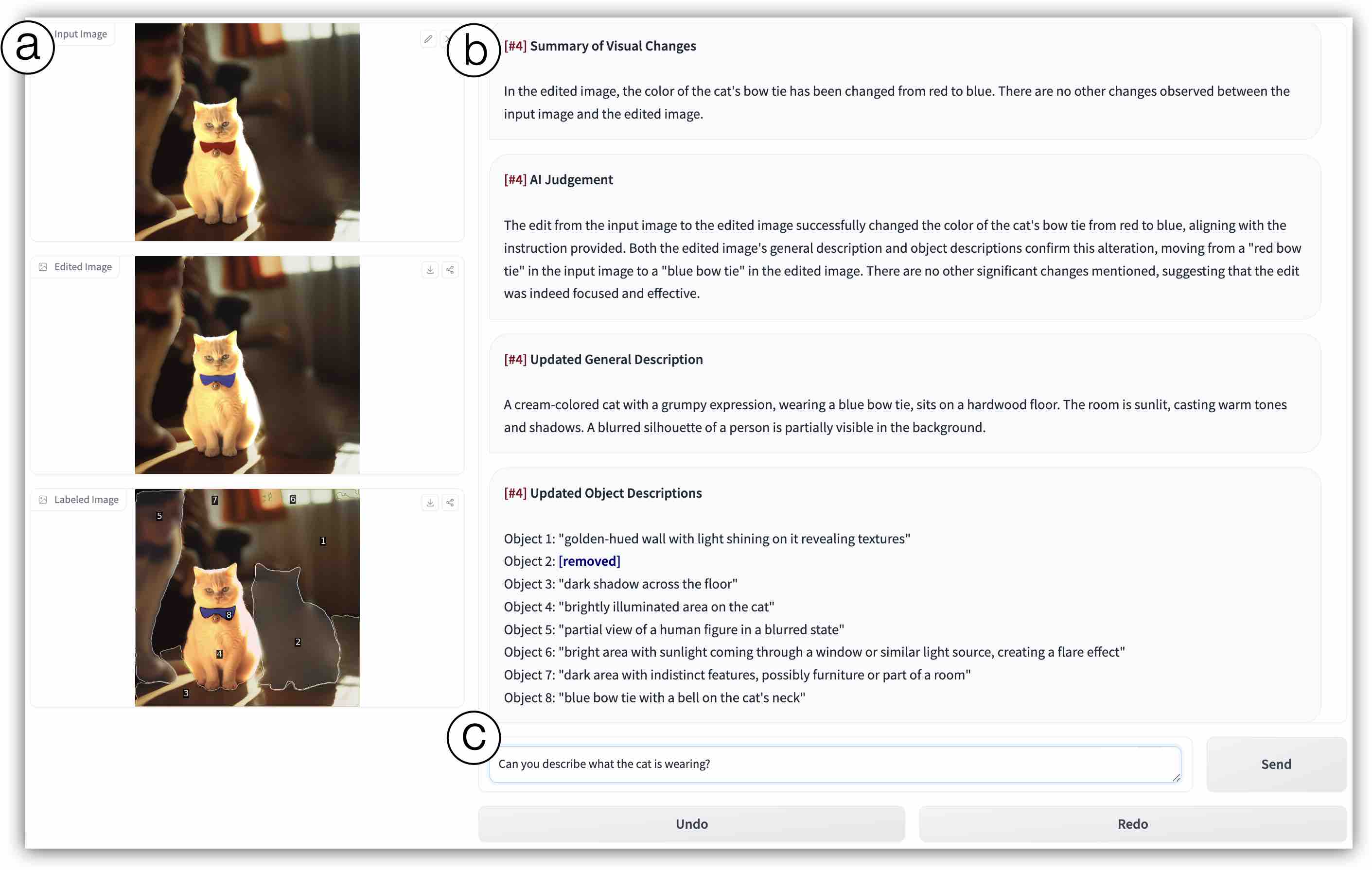}
\vspace{-1.pc}
\caption{
{\name} user interface. 
(a) The images before and after the most recent edit, and an image labeled with masks and indexes for debugging purposes.
(b) The accessible chat tagged with different heading levels helps users navigate a history of input prompts and verification feedback.
(c) Users can input natural language prompts to perform edits or ask follow-up questions, and undo or redo edits.
}
\vspace{-1pc}
\label{fig:app}
\Description{EditScribe user interface. 
(a) The images before and after the most recent edit, and an image labeled with masks and indexes for debugging purposes. The figure include the same images as in figure 1, with the first one showing the two cats, one is white with red bow tie and another orange one, and a woman kneeling on the floor.  The second one is the same image but the white cat’s bow tie is changed to blue. And the third images with segmentation masks and indexes labeled on.
(b) The accessible chat tagged with different heading levels helps users navigate a history of input queries and verification feedback. 
There are four types of feedback in the image:
[#4] Summary of Visual Changes
In the edited image, the color of the cat’s bow tie has been changed from red to blue. There are no other changes observed between the input image and the edited image.
[#4] AI Judgement 
The edit from the input image to the edited image successfully changed the color of the cat’s bow tie from red to blue, aligning with the instruction provided. Both the edited image’s general description and object descriptions confirm this alteration, moving from a “red bow tie” in the input image to a “blue bow tie” in the edited image. There are no other significant changes mentioned, suggesting that the edit was indeed focused and effective.
[#4] Updated General Description 
A cream-colored cat with a grumpy expression, wearing a blue bowtie, sits on a hardwood floor. The room is submit, casting warm tones and shadows. A blurred silhouette of a person is partially visible in the background. 
[#4] Updated Object Descriptions
Object 1: “golden-hued wall with light shining on it revealing textures”
Object 2: [removed]
Object 3: “dark shadow across the floor”
Object 4: “Brightly illuminated area on the cat”
Object 5: “partial view of a human figure in a blurred state”
Object 6: “bright area with sunlight coming through a window or similar light source, creating a flare effect”
Object 7: “dark are with indistinct features, possibly furniture or part of a room”
Object 8: “blue bow tie with a bell on the cat’s neck”
(c) Users can input natural language queries to perform edits or ask follow-up questions, and undo or redo edits. In the text box, there is a question the user is typing: “Can you describe what the is wearing?” There is a send button with the text “send” on it next to the textbox. There are undo and redo buttons with the text on them as well. 
}
\end{center}
\end{figure*}

\subsection{AI-assisted Image Editing}
Recent advances on computer vision, large language and multimodal models \cite{kirillov2023segment, wang2024yolov9, yu2023inpaint, goodfellow2014generative, radford2021learning, touvron2023llama, chatgpt} have introduced vast opportunity for easing image content authoring. Image editing automation now can perform not only basic photo adjustment, \textit{e.g., }color adjustment~\cite{runway-color, adobe-color}, but also advanced feature modifications and creative generation \cite{adobe, fatima2020, sheynin2023emu, parmar2023}. For example, development in object detection and segmentation algorithms~\cite{kirillov2023segment, wang2024yolov9} now supports precise image editing, such as foreground extraction, object removal, and inpainting~\cite{rother2004, yildirim2023instinpaint}. Generative models \cite{goodfellow2014generative, rombach2022highresolution, lipman2023flow} opened up possibilities for realistic generation of new images as well as manipulation of image attributes through the latent space \cite{yang2023}. The emergence of text-to-image diffusion models, such as DALL-E 2~\cite{ramesh2022hierarchical}, Stable Diffusion~\cite{rombach2022highresolution}, Promp-to-Prompt~\cite{hertz2022prompttoprompt}, and InstructPix2Pix~\cite{brooks2023instructpix2pix} supports not only text-to-image generation but also text-instructed edits on the image \cite{parmar2023, valevski2023, sheynin2023emu}.

So far, most AI-based image editing tools target at and have been tested with sighted image editors, primarily for efficiency- and inspiration-aid purposes. Only a limited number of accessibility research has explored how these tools may or may not support BLV individuals, a group whose content creation needs could critically benefit from this type of technology~\cite{lotus23assets, alharbi2022cscw}. Huh et al. prototyped a screen reader accessible text-to-image generation tool and evaluated with BLV creators~\cite{genassist}. Their study revealed design insights for accessible text-to-image \textit{generation} (\textit{e.g., }support for prompt generation and result image understanding). In terms of accessible image \textit{editing} support, research studies have focused primarily on how object detection, image caption, as well as segmentation models may be applied to support the BLV community’ visual privacy preservation---specifically, the detection and obfuscation of private and sensitive information in their photos~\cite{alharbi2022cscw, lotus23assets}. So far, this research revealed promising benefits these methods could bring to BLV individuals (\textit{e.g., }independence, sense of control, efficiency), but also challenges (\textit{e.g., }algorithm inaccuracies, difficulties with envisioning and evaluating obfuscation results, heavy cognitive load). 
Natural language interaction has the potential to mitigate these challenges by facilitating mutual understanding between the BLV users and the AI models. 
However, little insight exists around how BLV individuals would leverage this method to elicit feedback from AI models and how AI models could elaborate visual effects in the context of image editing.
To understand this loop of communication, we build a prototype system, {\name}, to explore how BLV individuals strategize prompts for the system to understand their editing needs, and how they perceive a set of verification feedback regarding the effects of their edits.

\begin{figure*}[t]
\vspace{-1pc}
\begin{center}
\includegraphics[width=\linewidth]{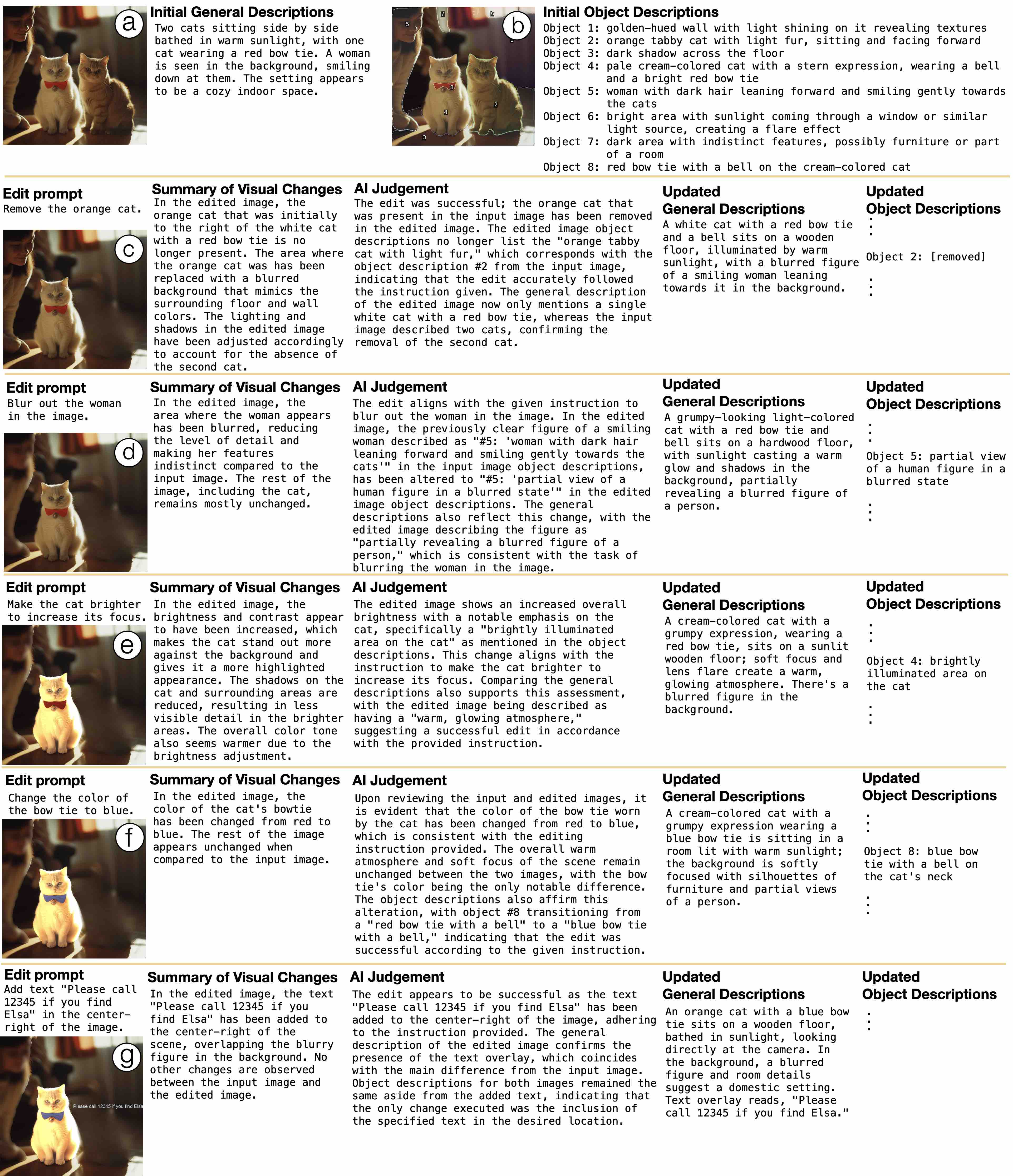}
\vspace{-1.5pc}
\caption{\CHANGE{Detailed edit prompts and corresponding verification feedback for the Walkthrough and Session 3 in our study. 
Note that we only show the updated object descriptions for edited objects due to space constraint. The user can access all object descriptions if needed on the {\name} interface, as demonstrated in Figure \ref{fig:app}. In (g), the object descriptions remained the same as before since the user did not specify an object to edit.}
}
\vspace{-2pc}
\label{fig:cat_scenario}
\Description{This figure demonstrate the verification output corresponding to certain edit query in session 3 and the walkthrough scenario. 
Here is the initial general descriptions of the image: Two cats sitting side by side bathed in warm sunlight, with one cat wearing a red bow tie. A woman is seen in the background, smiling down at them. The setting appears to be a cozy indoor space.
Here is the initial object descriptions of the image: 
Object 1: golden-hued wall with light shining on it revealing textures
Object 2: orange tabby cat with light fur, sitting and facing forward
Object 3: dark shadow across the floor
Object 4: pale cream-colored cat with a stern expression, wearing a bell and a bright red bow tie
Object 5: woman with dark hair leaning forward and smiling gently towards the cats
Object 6: bright area with sunlight coming through a window or similar light source, creating a flare effect
Object 7: dark area with indistinct features, possibly furniture or part of a room
Object 8: red bow tie with a bell on the cream-colored cat
Here is the verification feedback of the edit query 1: Remove the orange cat.
Summary of visual changes: In the edited image, the orange cat that was initially to the right of the white cat with a red bow tie is no longer present. The area where the orange cat was has been replaced with a blurred background that mimics the surrounding floor and wall colors. The lighting and shadows in the edited image have been adjusted accordingly to account for the absence of the second cat.
AI Judgement: The edit was successful; the orange cat that was present in the input image has been removed in the edited image. The edited image object descriptions no longer list the "orange tabby cat with light fur," which corresponds with the object description #2 from the input image, indicating that the edit accurately followed the instruction given. The general description of the edited image now only mentions a single white cat with a red bow tie, whereas the input image described two cats, confirming the removal of the second cat.
Updated general descriptions: A white cat with a red bow tie and a bell sits on a wooden floor, illuminated by warm sunlight, with a blurred figure of a smiling woman leaning towards it in the background.
Updated object descriptions: Object 2: [removed]
Here is the verification feedback of the edit query 2: Blur out the woman in the image.
Summary of visual changes: In the edited image, the area where the woman appears has been blurred, reducing the level of detail and making her features indistinct compared to the input image. The rest of the image, including the cat, remains mostly unchanged.
AI Judgement: The edit aligns with the given instruction to blur out the woman in the image. In the edited image, the previously clear figure of a smiling woman described as "#5: 'woman with dark hair leaning forward and smiling gently towards the cats'" in the input image object descriptions, has been altered to "#5: 'partial view of a human figure in a blurred state'" in the edited image object descriptions. The general descriptions also reflect this change, with the edited image describing the figure as "partially revealing a blurred figure of a person," which is consistent with the task of blurring the woman in the image.
Updated general descriptions: A grumpy-looking light-colored cat with a red bow tie and bell sits on a hardwood floor, with sunlight casting a warm glow and shadows in the background, partially revealing a blurred figure of a person.
Updated object descriptions: Object 5: partial view of a human figure in a blurred state
Here is the verification feedback of the edit query 3: Make the cat brighter to increase its focus.
Summary of visual changes:In the edited image, the brightness and contrast appear to have been increased, which makes the cat stand out more against the background and gives it a more highlighted appearance. The shadows on the cat and surrounding areas are reduced, resulting in less visible detail in the brighter areas. The overall color tone also seems warmer due to the brightness adjustment.
AI Judgement: The edited image shows an increased overall brightness with a notable emphasis on the cat, specifically a "brightly illuminated area on the cat" as mentioned in the object descriptions. This change aligns with the instruction to make the cat brighter to increase its focus. Comparing the general descriptions also supports this assessment, with the edited image being described as having a "warm, glowing atmosphere," suggesting a successful edit in accordance with the provided instruction.
Updated general descriptions: A cream-colored cat with a grumpy expression, wearing a red bow tie, sits on a sunlit wooden floor; soft focus and lens flare create a warm, glowing atmosphere. There's a blurred figure in the background.
Updated object descriptions: Object 4: brightly illuminated area on the cat
Here is the verification feedback of the edit query 4: Change the color of the bow tie to blue.
Summary of visual changes: In the edited image, the color of the cat's bowtie has been changed from red to blue. The rest of the image appears unchanged when compared to the input image.
AI Judgement: Upon reviewing the input and edited images, it is evident that the color of the bow tie worn by the cat has been changed from red to blue, which is consistent with the editing instruction provided. The overall warm atmosphere and soft focus of the scene remain unchanged between the two images, with the bow tie's color being the only notable difference. The object descriptions also affirm this alteration, with object #8 transitioning from a "red bow tie with a bell" to a "blue bow tie with a bell," indicating that the edit was successful according to the given instruction.
Updated general descriptions: A cream-colored cat with a grumpy expression wearing a blue bow tie is sitting in a room lit with warm sunlight; the background is softly focused with silhouettes of furniture and partial views of a person.
Updated object descriptions: Object 8: blue bow tie with a bell on the cat's neck
Here is the verification feedback of the edit query 5: Add text "Please call 12345 if you find Elsa" in the center-right of the image.
Summary of visual changes: In the edited image, the text "Please call 12345 if you find Elsa" has been added to the center-right of the scene, overlapping the blurry figure in the background. No other changes are observed between the input image and the edited image.
AI Judgement: The edit appears to be successful as the text "Please call 12345 if you find Elsa" has been added to the center-right of the image, adhering to the instruction provided. The general description of the edited image confirms the presence of the text overlay, which coincides with the main difference from the input image. Object descriptions for both images remained the same aside from the added text, indicating that the only change executed was the inclusion of the specified text in the desired location.
Updated general descriptions: An orange cat with a blue bow tie sits on a wooden floor, bathed in sunlight, looking directly at the camera. In the background, a blurred figure and room details suggest a domestic setting. Text overlay reads, "Please call 12345 if you find Elsa."
Updated object descriptions: None
}
\end{center}
\end{figure*}

\section{{\name}}
{\name} leverages natural language verification loops as its core to support BLV people to understand and edit images non-visually.
In this section, we first present how BLV users could use {\name} to edit images in an example scenario, then present the inspiration to design natural language verification loops from prior work, and detail its implementation.

\subsection{Scenario Walkthrough}\label{walkthrough}
Here, we illustrate {\name} in an example scenario, taking Amanda as the main character, a person who is blind.

Amanda just lost her short-haired British white cat, Elsa, and urgently wants to make a public post asking people to help find it. 
She navigates her album on the phone and found the latest photo of herself with her cats, Elsa and Rosa. 
She uploads this photo to {\name} and obtains a general description and a list of detailed object descriptions (Figure \ref{fig:cat_scenario}a,b) to confirm that the image content matches what she remembered. 
With these descriptions, she has developed a plan in mind to make a post with this photo.

First, Amanda decides to remove Rosa, her other orange tabby cat (Figure \ref{fig:cat_scenario}c) next to Elsa, to avoid confusion. 
She prompts \PROMPT{remove the orange cat} and received {\name} feedback after the edit is completed, including \textsc{Summary of Visual Changes}, \textsc{AI Judgement}, \textsc{Updated General Descriptions}, and \textsc{Updated Object Descriptions}. 
Amanda confirms Rosa has been removed by cross-checking {\name}'s feedback, \textit{e.g., }\textsc{Summary of Visual Changes} confirms the edit and \textsc{AI Judgement} provides detailed reasons, only one cat is described in the \textsc{Updated General Descriptions} and the removal of Rosa's description in \textsc{Updated Object Descriptions} (Figure \ref{fig:cat_scenario}c). 
She further confirms with a question \PROMPT{How many cats are in the image?} with answer \PROMPT{One.}
Amanda then blurs herself to maintain her privacy by prompting \PROMPT{Blur out the woman in the image}, and also confirms the success of the edit through the consistent feedback from {\name} (Figure \ref{fig:cat_scenario}d). 

To help people better distinguish Elsa, she wants to highlight it and make the photo similar to when Elsa was lost. 
First, Amanda aims to increase the focus of Elsa and specifies \PROMPT{Make the cat brighter to increase its focus.} 
The feedback from {\name} suggests \PROMPT{the intensity of the sunlight on the cat has been increased, giving the cat a more illuminated appearance} (Figure \ref{fig:cat_scenario}e).
She wants to ensure the color of Elsa is still maintained and asks \PROMPT{What is the color of the cat?} with the answer \PROMPT{Cream or white}, which aligns with how her friends describe Elsa before.
Next, she changes the bow tie's color from red to blue, which is what Elsa wore when lost \PROMPT{Change the color of the bow tie to blue.} She confirms with {\name}'s feedback, especially the \textsc{updated object description} indicates \PROMPT{blue bow tie with a bell on the cat's neck} (Figure \ref{fig:cat_scenario}f).

Finally, Amanda wants to add her phone number to the post for others' reference. 
She first asks {\name} \PROMPT{Dividing the image to nine squares, where is a good spot to add text that does not overlap the cat?} with {\name} answering \PROMPT{Center Right.} She then prompts {\name} \PROMPT{add text "Please call 12345 if you find Elsa" in the center-right of the image} (Figure \ref{fig:cat_scenario}g). 
Despite the {\name} feedback confirming the addition of text, she performs another check \PROMPT{Does the text overlap with the cat?} with answer \PROMPT{No}, and another check \PROMPT{Does the cat with blue bow tie stand out in the image?} with answer \PROMPT{Yes.}
Finally, she confidently posts it on her social media and waits for responses. 

\subsection{Natural Language Verification Loops}
When sighted people perform image editing, they utilize visual feedback to verify their actions and make necessary adjustments.
However, such visual verification loops are inaccessible to BLV people.
Therefore, at its core, {\name} is enabled by the idea of \textit{natural language verification loops} to support non-visual image editing. 
Here, we introduce its high-level concept.

LMMs (\textit{e.g., }GPT-4v~\cite{gpt4v}) are leveraged in {\name} as a proxy to communicate between the user and the image. 
{\name} interprets the user's natural language prompt into the action of certain visual edits, and then converts the resulting visual changes back into textual feedback for the user to review and confirm.
This loop of cross-modal communication can be complex and requires extensive textual exchanges.
First, from applying textual prompts to visual changes, {\name} interprets the object or region of the user's interest and desired editing effect by grounding on the user's prompt and the image content. 
Second, from interpreting visual changes to textual feedback, {\name} grounds on cross-modal sources to provide a set of verification feedback, which we describe their rationale:
\begin{enumerate}
\item \textbf{\textsc{Summary of Visual Changes}} is designed to resemble and simulate a sighted person editing an image, who can visually discern the main differences between the images before and after edits.
\item \textbf{\textsc{AI Judgement}} draws inspiration from a human judge, who, from a second-person perspective, assesses both visual and textual modifications to evaluate and rationalize the success of the edit.
\item \textbf{\textsc{Updated General Descriptions}} is designed to offer an independent perspective on how a sighted person would perceive the new image after an edit~\cite{wu2017, diagramcenter2015, w3coverview}.
\item \textbf{\textsc{Updated Object Descriptions}} is designed as if a sighted person is taking detailed visual inspections of each object to discern the nuances. 
\end{enumerate}
The updated general and object descriptions are also inspired by previous research on presenting image descriptions with varying levels of granularity to minimize cognitive load and enhance efficiency~\cite{lee2022imageexplorer, stangl21assets, van2024making, avscript}.
\CHANGE{The provision of four types of feedback, each grounded in different sources, aims to help mitigate the potential AI errors (\textit{e.g.,} misinterpreting user intent or generating hallucinations.) by enabling users to compare the feedback and judge their accuracy.
}
Below, we detail the cross-modal grounding pipeline for understanding user prompts and generating verification feedback.

\subsection{Cross-modal Grounding Pipeline for Understanding User Prompts and Generating Verification Feedback}\label{pipeline}
To facilitate image content understanding, {\name} first generates initial general and object descriptions for the user. 
Then, the user can perform edits with natural language prompts, and get verification feedback after each edit. 
The user can ask follow-up questions to clarify the verification feedback and verify the edits. 
They can repeat this process until the edited image satisfies their goal.

\textit{\textbf{Generating General and Object Descriptions.}} Given an input image, {\name} generates general and object descriptions to support the initial understanding of image content. 
Specifically, {\name} provides general descriptions for the image by prompting GPT-4v~\cite{gpt4v}. 
Next, {\name} generates object descriptions by Set-of-Mark Prompting \cite{setofmark} with input image overlaid with a visible bounding mask (by SEEM \cite{seem}) and index on each object:

\begin{figure}[h]
\vspace{-1pc}
\begin{center}
\includegraphics[width=\linewidth]{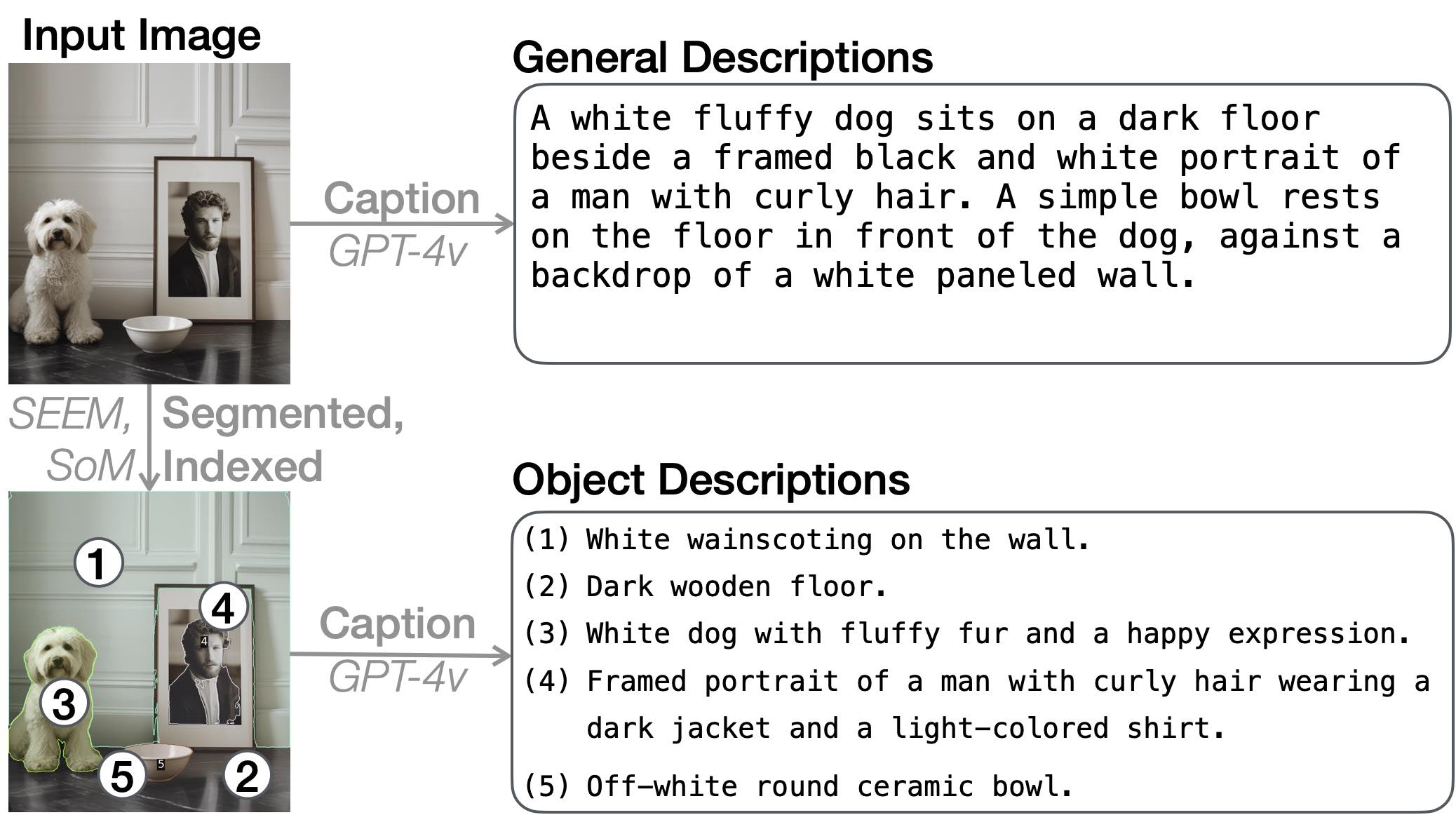}
\vspace{-1.5pc}
\label{fig:initial_descriptions}
\caption{Pipeline to generate initial general and object descriptions. {\name} prompts GPT-4v~\cite{gpt4v} with input image to generate general descriptions, and uses visual bounding masks (by SEEM \cite{seem}) and object indexes to perform Set-of-Mark Prompting to generate object descriptions. 
}
\Description{The figure shows a input image go through GPT-4v to generate general descriptions “A white fluffy dog sits on a dark floor beside a framed black and white portrait of a man with curly hair. A simple bowl rests on the floor in front of the dog, against a backdrop of a white paneled wall.”
The input image then goes through the segmentation and indexing process by models SEEM and SoM and generates detailed object descriptions like (1) white wainscoting on the wall. (2) dark wooden floor. (3) White dog with fluffy fur and da happy expression. (4) FRamed portrait of a man with curly hair wearing a dark jacket and a light-colored shirt. (5) Off-white round ceramic bowl.
}
\end{center}
\end{figure}

\textit{\textbf{Mapping User Prompt to Edit Action or Followup Question.}}
When receiving a user prompt, {\name} classifies it as either a \textit{question} or an \textit{edit instruction} that corresponds to one of the {\name} edit functions by prompting GPT-4.
If the prompt is classified as a \textit{question}, {\name} prompts the user's prompt to GPT-4v~\cite{gpt4v} to answer the question. 
On the other hand, if the prompt is classified as an \textit{edit instruction}, {\name} extracts the intended edit action and the object of interest from the prompt, for which the user can specify either the object's name or index for flexible referencing. Having all this information, {\name} performs the edit to the referenced object (See Section \ref{edit_actions} for the supported edit actions).

\begin{figure}[b]
\vspace{-0.5pc}
\begin{center}
\includegraphics[width=\linewidth]{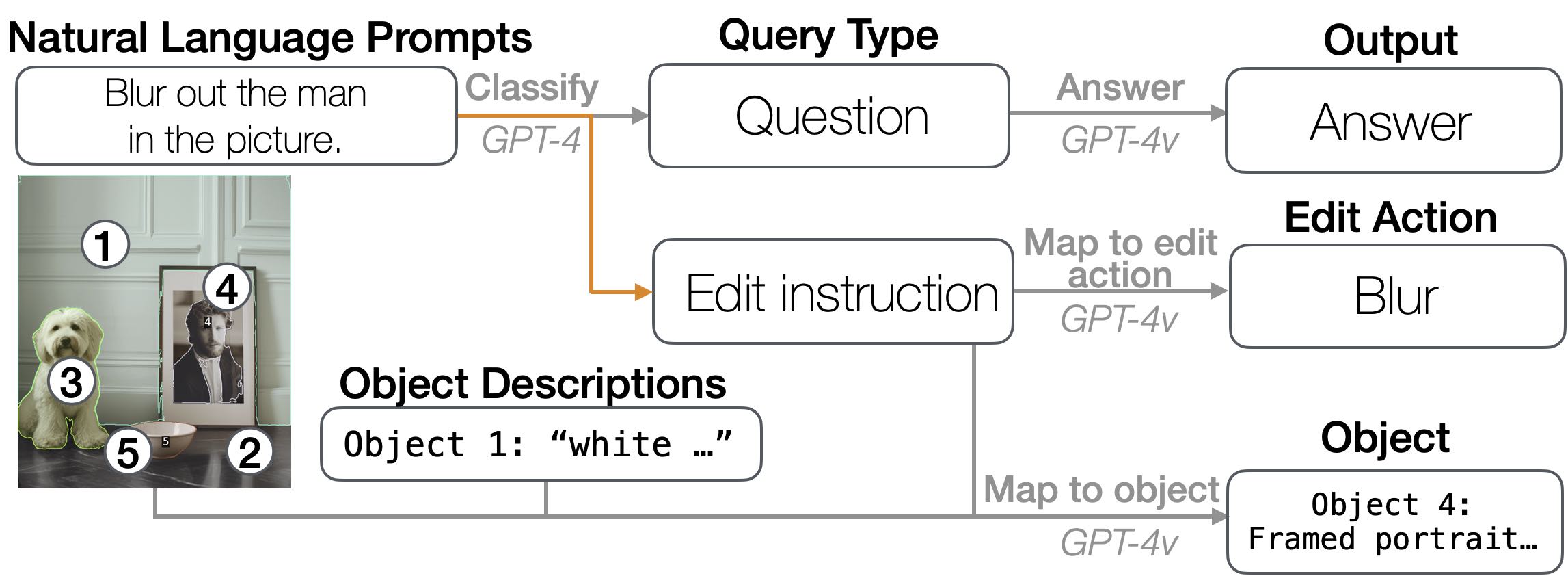}
\vspace{-1.5pc}
\label{fig:classify}
\caption{Pipeline to classify user prompts to actionable items. 
If the prompt is classified as a \textit{question}, {\name} prompts the prompt to GPT-4v~\cite{gpt4v} to answer the question.
if the prompt is classified as an \textit{edit instruction}, {\name} extracts the intended edit action and the object of interest from the prompt.
}
\Description{The figure shows the indexed image from Figure 4. There is a natural language prompts “blur out the man in the picture” and this prompt will be classified to either a question or an edit instruction. If it’s a question, then it will send to GPT-4v to process the answer. If it’s a edit instruction, then it will be mapped to a certain edit action, such as “blur” in this case, and it will be also mapped to a object of interest like Object 4: framed portrait… based on the object descriptions generated from previous step “Object 1 : white… ”. }
\end{center}
\end{figure}

\textit{\textbf{Verification Feedback.}}
After an edit is completed, {\name} generates four types of verification feedback for the user to verify the outcome of the edit, including \textsc{Summary of Visual Changes}, \textsc{AI Judgement}, \textsc{Updated General Descriptions}, and \textsc{Updated Object Descriptions}.
These verification information are structured and presented to the user from high to low level as follows:

\begin{enumerate}
    \item \textbf{\textsc{Summary of Visual Changes}} illustrates an overview of visual changes by comparing the images before and after the edit. It grounds only new and previous images for the prompt.

    \item \textbf{\textsc{AI Judgement}} explicitly points out if the AI determines the edit is successful or not by grounding and comparing the images before and after the edit, new and previous general and object descriptions, and the edit actions altogether. It also details its reasoning process and identified evidence. 

    \item \textbf{\textsc{Updated General Descriptions}} are new and independent general descriptions on only the edited image using the same prompt for generating the initial descriptions. It only takes the new image as input for the prompt.

    \item \textbf{\textsc{Updated Object Descriptions}} are object descriptions on the edited image, using SoM prompting \cite{setofmark}. It takes the new image with labeled masks and indexes for the prompt.

\end{enumerate}
\begin{figure}[h]
\begin{center}
\includegraphics[width=\linewidth]{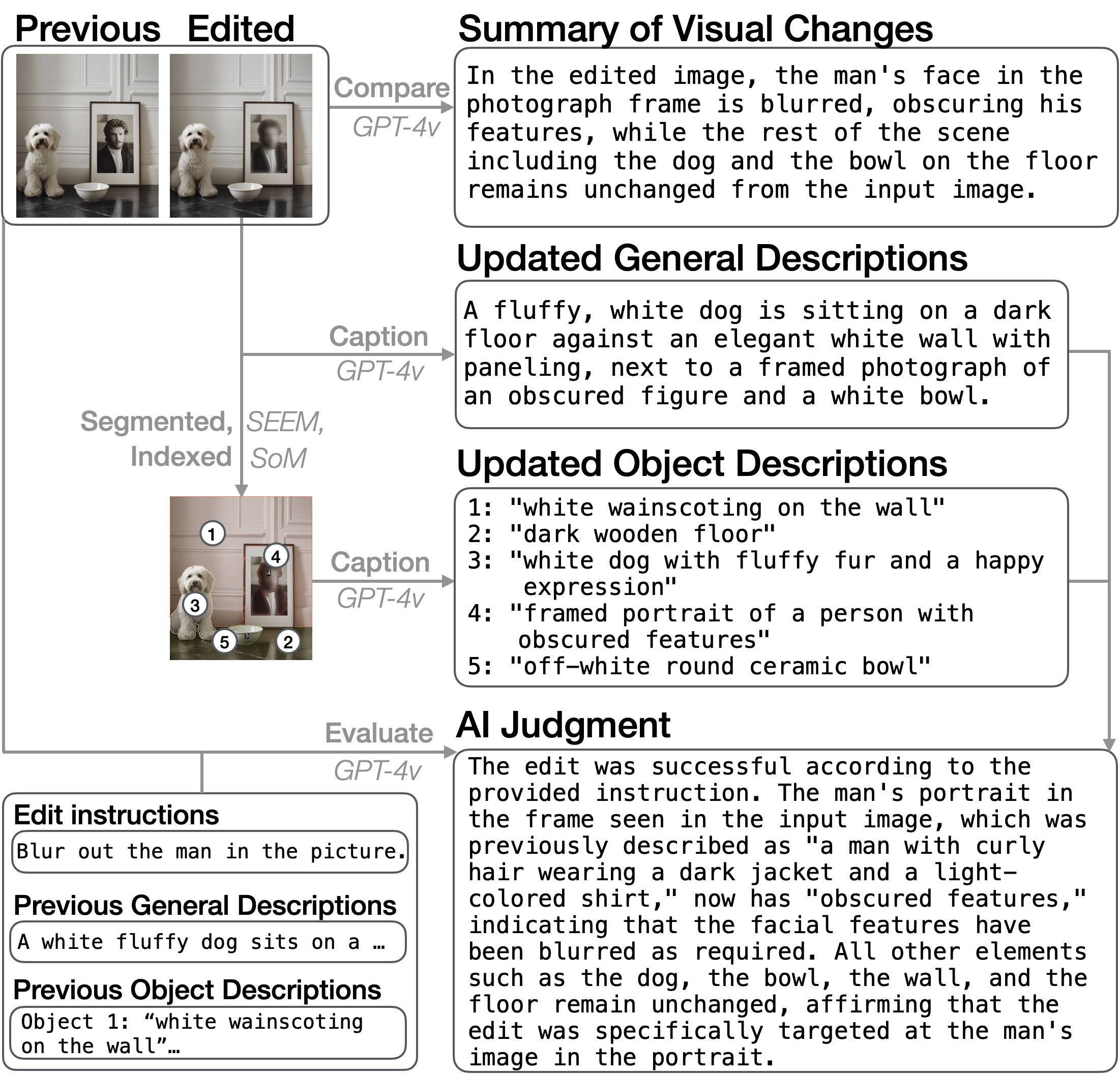}
\vspace{-1.5pc}
\label{fig:outputs}
\caption{Pipeline to generate the four types of verification feedback. \textit{Summary of Visual changes} compares the previous and edited images, while the \textit{updated general and object descriptions} takes only edited images as input for prompting. \textit{AI Judgment} takes both previous and edited images as input, as well as the texts, such as user prompts and previous and current general and object descriptions.
}
\Description{Same images of Figure 4,5 are shown in this figure. There is a previous image with no edit, and another edited image with the man in the picture blurred. The two images are sent to GPT-4v to compare and generate the summary of visual changes “In the edited image, the mna’s face in the photograph frame is blurred, obscuring his features, while the rest of the scene including the dog and the bowl on the floor remains unchanged from the input image.” At the same time, the edited image sent to GPT-4v to generate updated general descriptions “A fluffy, white dog is sitting on a dark floor against an elegant white wall with paneling, next to a framed photograph of an obscured figure and a white bowl”. The edited image is also segmented and indexed as an input image to GPT-4v to generate updated object descriptions “(1) white wainscoting on the wall (2) dark wooden floor (3) white dog with fluffy fur and a happy expression (4) framed portrait of a person with obscured features (5) off-white round ceramic bowl.” The system then take all of these information and edit instructions “blur out the man in the picture”, as well as previous general and object descriptions to generate AI judgement “The edit was successful according to the provided instruction. The man's portrait in the frame seen in the input image, which was previously described as "a man with curly hair wearing a dark jacket and a light-colored shirt," now has "obscured features," indicating that the facial features have been blurred as required. All other elements such as the dog, the bowl, the wall, and the floor remain unchanged, affirming that the edit was specifically targeted at the man's image in the portrait.”}
\end{center}
\end{figure}

\subsection{Image Edit Actions}\label{edit_actions}
{\name} supports five editing functions: \textit{(i) blur an object}, \textit{(ii) remove an object}, \textit{(iii) change color of an object}, \textit{(iv) adjust brightness of an object}, and \textit{(v) add text to the image}. 
\CHANGE{
We focus on object-level actions as they are critical to tasks commonly desired by BLV people \cite{alharbi2022cscw, bennett2018, voykinska2016}, \textit{e.g.,} image obfuscation for privacy \cite{alharbi2022cscw, lotusChi24} and background removal/blurring to focus on specific objects \cite{bennett2018, visphoto}.
These actions require precise understanding and manipulation of image details, and we selected a subset of them that are representative and could inform other tasks. For instance, object removal could inform replacing with new objects, or cropping an image. Blurring and changing color and brightness could inform other actions involving pixel modifications. Inserting texts could inform adding other visual elements such as icons or arrows. 
}
We can further augment editing functions in {\name} by prompting LLMs to classify users' prompts to corresponding edit functions.
Below, we exemplify how users could specify prompts for each edit action using natural language, but not limited to these phrasing:

\begin{enumerate}[noitemsep, leftmargin=*]
    \item \textit{Blur an object}: Blurring is a common technique to preserve personal image content privacy \cite{alharbi2022cscw,lotusChi24}. Users can specify prompts relevant to blurring, such as \PROMPT{blur the person out}, or \PROMPT{make \#2 vague.} Based on the prompt, {\name} will apply the blurry effect on the specified object, using OpenCV's built-in function `GaussianBlur' \cite{opencv}.
    Users can perform this edit action repeatedly on an object until its level of blurriness meets their expectations.

    \item \textit{Remove an object}: Removing unwanted or personal image content is also common in image editing tasks. With this action, the user can remove an object while preserving the background. Users can specify prompts such as \PROMPT{remove the right person}, or \PROMPT{remove the orange cat.} {\name} will then remove the specified object using LaMa \cite{suvorov2021resolution}.

    \item \textit{Change color of an object}: Changing color is common in basic image editing apps for improving aesthetics (e.g.,~\cite{adobe-color}). It is also a privacy-preserving technique for users to black out or fill an object with a color significantly different from its original appearance~\cite{alharbi2022cscw,lotusChi24}. 
    Users can specify prompts, such as \PROMPT{change the cat's collar to blue.} 
    {\name} will then change the color of the specified object to the specified one.
    This is achieved by modifying the `Hue' value of the object in the HSV color space (\textit{e.g.,} Hue, Saturation, Value), which represents the color type and is expressed as a degree on the color wheel, ranging from 0 to 360. 
    
    \item \textit{Adjust the brightness of an object}: Adjusting the brightness is another common editing function. Users can input prompts such as \PROMPT{increase the brightness of the \#6 person}, or \PROMPT{make the left cat brighter.}
    {\name} will then adjust the brightness by increasing or decreasing each RGB channel of a pixel with the same increments.
    Users can gradually adjust an object's brightness until its brightness level meets their expectations.

    \item \textit{Add text to the image}: Adding text is common for adding an explanation to an image or making any purpose explicit. In our current implementation, users can assign text to one of the nine locations on an image (\textit{e.g., }center, bottom right) or onto a specific object. Users can specify \PROMPT{add words `Hello world' on upper third}, \PROMPT{place sentence `Hello world' to center}, or \PROMPT{generate text `Hello' to the top left corner.} {\name} will then place the text to the specified location using OpenCV's built-in function `addText' \cite{opencv} with pre-defined colors based on the contrast to the image.
\end{enumerate}

\subsection{{\name} Web Interface}
{\name} was implemented using Gradio \cite{gradio}, an open-source Python library for fast prototyping front-end web interfaces.
The left side of the interface (Figure \ref{fig:app}a) shows the images before and after the most recent edit, and an image labeled with masks and indexes for debugging purposes. 
On the right side (Figure \ref{fig:app}b), there is an accessible chat with verification feedback indexed with different heading levels for BLV users to navigate using a screen reader. 
\CHANGE{For instance, verification feedback of each edit starts with ``Verification Output of Edit \#4 starts from here'', which is labeled as heading level 1 to help BVI users navigate the edits. On the other hand, each type of feedback title, such as ``[\#4] Summary of Visual Changes'', is labeled as heading level 2, which allows users to navigate each type of feedback.}
On the bottom right of the interface (Figure \ref{fig:app}c), the user can enter their prompts and questions using natural language, or ``undo'' or ``redo'' their edits.  

\begin{table*}[t]
  \caption{\CHANGE{Participant demographics information.}}
  \label{tab:demographic}
  \vspace{-1pc}
  \begin{center}
  \begin{tabular}{|l|l|l|p{2.0cm}|p{1.5cm}|p{3.5cm}|p{5.5cm}|}
    \hline
    \textbf{PID} & \textbf{Age} & \textbf{Gender} & \textbf{Vision Level} & \textbf{BLV Onset} & \textbf{Experience with GenAI} & \textbf{Experience in image creation, editing or verification} \\
    \hline
    P1 & 43 & Male & Low vision & Since 30yo & Using daily & Adding basic existing graphics, \textit{e.g.,} arrows, circles, to existing images \\
    \hline
    P2 & 42 & Male & Total Blindness & Since 9yo & Using daily & None \\
    \hline
    P3 & 40 & Male & Total blindness & Since 2yo & & Cropping or rotating images  \\
    \hline
    P4 & 40 & Female & Total blindness & Birth & & None \\
    \hline
    P5 & 43 & Male & Some light perception & Birth & Using daily. & Resizing or flipping images \\
    \hline 
    P6 & 29 & Male & Total blindness & Birth & Creating and iterating images with Midjourney & Using existing captioning apps or tactile display for verifying content \\
    \hline
    P7 & 30 & Female & Total blindness & Since 17yo & Using GPT for work & Python or R programming to create or edit graphs for work purposes \\
    \hline
    P8 & 34 & Male & Total Blindness & Since 18yo & & None \\
    \hline
    P9 & 26 & Male & Total Blindness & Since 16yo &  Creating and iterating images with Midjourney and ChatGPT & A photographer. Using the iOS photo app for basic tasks like exposure, applying filters, contrast, and rotating the image \\
    \hline
    P10 & 27  & Male & Total blindness & Birth &  & None  \\
    \hline
  \end{tabular}
  \end{center}
  \Description{}
\end{table*}

\section{User Study}
We conducted a user study with BLV individuals to understand how natural language verification loops, including user prompts and system verification feedback, may (or may not) support their image editing needs. Specifically, we focus on the following research questions: 
\begin{enumerate}
    \item[RQ1:] How does {\name} support non-visual image editing?
    \item[RQ2:] How do BLV people prompt {\name}?
    \item[RQ3:] How do BLV people perceive {\name}'s verification\\ feedback?
    \item[RQ4:] How do BLV people perceive the final edited images?
\end{enumerate}

\subsection{Participants}
We recruited ten BLV participants (8 male and 2 female) using mailing lists for local accessibility organizations and prior contacts.
Participants aged from 26 to 43 (avg. 35.4) and described their visual impairment as total blindness (N=8), having only light perception (N=1) or low vision (N=1).
All participants regularly used a screen reader as the primary means to accessing their devices.
Some participants had prior experiences in taking photos or editing images, such as having attempted to crop images, adding text or graphics, creating images with Generative AI, or consuming images through tactile displays (Table \ref{tab:demographic}).

\subsection{Study Procedure and Sessions}
\begin{figure}[t]
\begin{center}
\includegraphics[width=\linewidth]{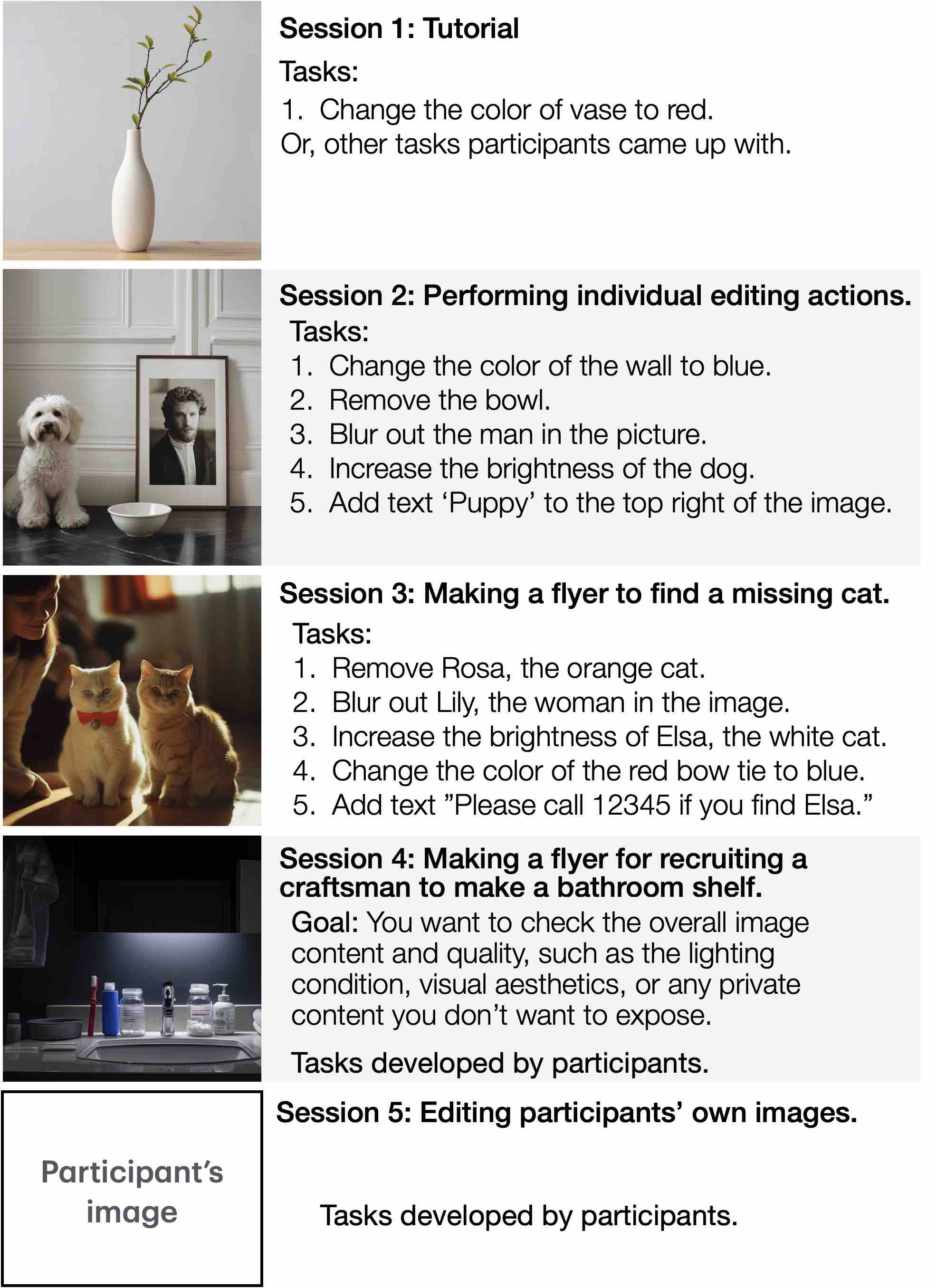}
\vspace{-1.5pc}
\caption{
Images and tasks used in the user study sessions. Please refer to Figure \ref{fig:cat_scenario}, \ref{fig:dog_scenario}, and \ref{fig:private_scenario} for more details.
}
\vspace{-1pc}
\label{fig:tasks}
\Description{Session 1: tutorial. 
Tasks: Change the color of vase to red.
Or, other tasks participants came up with.
Image description: A single white vase with a narrow neck sits on a wooden surface, containing a branch with budding green leaves. The background is neutral and plain.
Session 2: Performing individual editing actions.
Tasks
1. Change the color of the wall to blue.
2. Remove the bowl.
3. Blur out the man in the picture.
4. Increase the brightness of the dog.
5. Add text ‘Puppy’ to the top right of the image.
Image descriptions: A white fluffy dog sits on a dark floor beside a framed black and white portrait of a man with curly hair. A simple bowl rests on the floor in front of the dog, against a backdrop of a white paneled wall.
Session 3: Making a flyer to find a missing cat.
Tasks:
1. Remove Rosa, the orange cat.
2. Blur out Lily, the woman in the image.
3. Increase the brightness of Elsa, the white cat.
4. Change the color of the red bow tie to blue.
5. Add text ”Please call 12345 if you find Elsa.”
Image description: Two cats sitting side by side bathed in warm sunlight, with one cat wearing a red bow tie. A woman is seen in the background, smiling down at them. The setting appears to be a cozy indoor space.
Session 4: Making a flyer for recruiting a craftsman to make a bathroom shelf.
Goal: You want to check the overall image content and quality, such as the lighting condition, visual aesthetics, or any private content you don’t want to expose. Tasks developed by participants.
Session 5: Editing participants’ own images. Tasks developed by participants.
}
\end{center}
\end{figure}

The study was conducted remotely over Zoom, and participants were provided with a link to our {\name} site and used a screen reader to access. The study was approved by our Institutional Review Board (IRB).
With participants' consent, the study was video recorded and took about two hours, and each participant was compensated \$50 for their participation. 
Participants were asked to optionally provide one of their own images to edit in the study. 
The study had five sessions, including a tutorial session to familiarize participants with {\name}, three sessions to edit our provided images, and a session to edit their provided images (Figure \ref{fig:tasks}).
We developed several tasks in each session, which we describe next.

\subsubsection{Session 1: Tutorial.}
In this session, we aimed to familiarize our participants with {\name}. 
Participants were guided through each element on the {\name} UI and were asked to perform editing tasks of their interests on an example image. 
Participants were instructed on the {\name} feedback after an edit, including the four types of outputs, and noted that they could ask follow-up questions to verify the image content as well as undo or redo their edits (Figure \ref{fig:app}). 

\subsubsection{Session 2: Performing individual edit actions.}
In this session, we aimed to understand if participants could successfully perform each edit action of {\name}.
For each edit action, participants were asked to use natural language prompts to edit or verify edits until they confirmed the task was completed. 
We then asked participants to provide their agreement on the statement \textit{``I am confident that the edit is successful''} on a scale from 1 to 7, where 1 is strongly disagree and 7 is strongly agree. 
We randomized the order of the five editing tasks across participants.

\subsubsection{Session 3: Making a flyer to find a missing cat.}
In this session, we aimed to explore how participants would use {\name} in a practical scenario. Participants engaged in the scenario detailed in Section \ref{walkthrough}.
Participants were asked to perform the tasks in their preferred order until they confirmed or were satisfied with the results, or if they decided to move on.

\subsubsection{Session 4: Making a flyer for recruiting a craftsman to make a bathroom shelf.} 
Next, instead of providing individual specific tasks, we provided a high-level goal for participants to develop their editing tasks based on their needs.
The image was more complex, with multiple objects and uneven lighting conditions (Figure \ref{fig:tasks}).
The instruction was:
\textit{
You plan to create a flyer and post it on your social media to find someone to custom-make a shelf to fit everything on your sink. 
You thus first took a photo of the sink in your bathroom. 
However, it includes many personal items, such as towels, containers, medication bottles, toothbrushes. 
You want to check the overall image content and quality, such as the lighting condition, visual aesthetics, or any private content you don’t want to expose. 
You should develop three to five edit actions. 
}

\subsubsection{Session 5: Editing participants' own images.}
In this session, we aimed to encourage participants to brainstorm potential use cases and workflows to apply {\name} on their own images.
Five participants provided their images before the study, for which they were familiar with the content and context when the photos were taken. 
In this open-ended session, participants performed their goals without any restrictions. 

\subsection{Data Analysis}
Besides reporting their perceived confidence in each editing task in session 2, 
participants were asked to think aloud during each session and interviewed afterwards to provide qualitative feedback on their strategies for creating prompts or perceiving {\name} outputs.
All sessions were recorded and transcribed for analysis. The first author independently performed open coding on all transcripts to identify initial codes. The initial coding focused on participants' prompt formation and decision making, feedback toward the system output, and any friction they experienced. Two authors reviewed all coded excerpts and iterated on the set of codes through discussing with the first author. 
They then conducted affinity diagramming~\cite{hartson2012ux} on the initial codes to extract and organize high-level themes.

\begin{figure*}[t]
\begin{center}
\includegraphics[width=\linewidth]{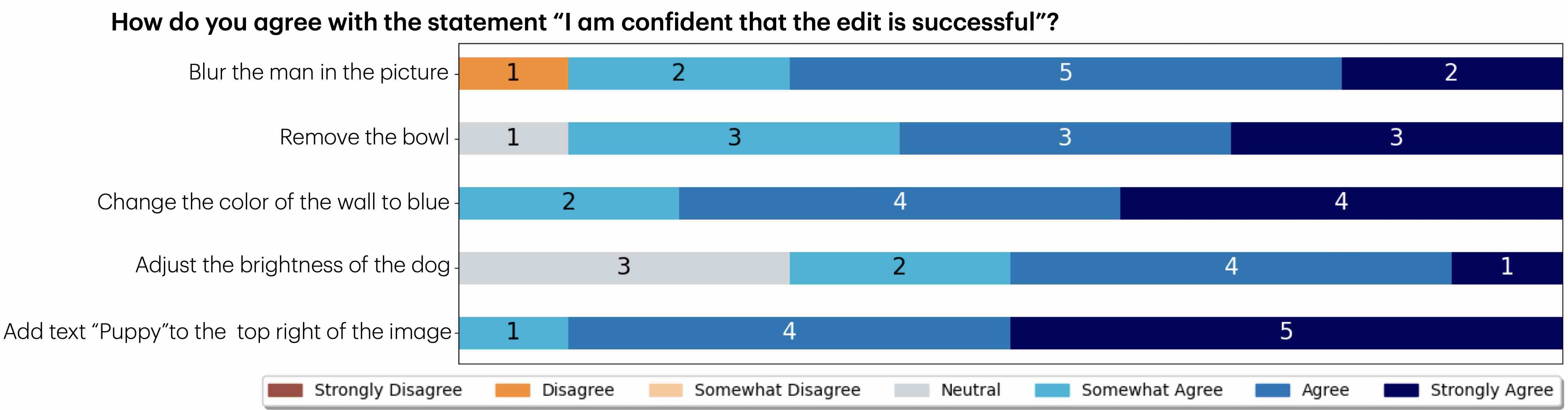}
\vspace{-1.5pc}
\caption{Likert scale responses of participants' perceived confidence for each edit action in study session 2.  
}
\label{fig:results}
\Description{Likert scale responses of participants' perceived confidence for each editing action in study session 2. 
How do you agree with the statement “I am confident that the edit is successful”?
For the task of blur the man in the picture, one disagree, two somewhat agree, five agree, two strongly agree.
For the task of remove the bowl, one neutral, three somewhat agree, three agree, three strongly agree
For the task of change the color of the wall to blue, two somewhat agree, four agree, four strongly agree
For the task of adjust the brightness of the dog, three neutral, two somewhat agree, four agree, one strongly agree
For the task of add text “puppy” to the top right of the image, one somewhat agree, four agree, five strongly agree
}
\end{center}
\end{figure*}

\section{Findings}

\subsection{RQ1: How does {\name} support non-visual image editing?}\label{RQ1_results}
\textit{Participants were able to successfully complete most of the tasks in the study sessions and found {\name} promising in supporting everyday scenarios. On the other hand, participants also expressed a need for more edit actions and finer control over image editing.}

\subsubsection{Task performance in the study sessions.}\label{R_task_completion}
In session 2, all participants were able to complete all the tasks (Figure \ref{fig:tasks}) through natural language verification loops, but with varying confidence about the editing results. 
In particular, 
they were generally confident about the results of changing the color of the wall (\(\mu\)=6.2, \(\sigma\)=0.8) and adding text to the image (\(\mu\)=6.4, \(\sigma\)=0.7), which was generally straightforward and consistent across verification feedback. 
However, there was occasional confusing verification feedback in the other three tasks, which made participants less confident about the results. 
First, for the task of blurring out the person in the picture (\(\mu\)=5.6, \(\sigma\)=1.4), 
P8, who scored 2 for this task, was concerned if only the face were blurred but not the entire person \QUOTE{I would be lower on this, so I gave two. It says the face was blurred, but I expected it to blur out the entire body.} 
For removing the bowl (\(\mu\)=5.8, \(\sigma\)=1.0), P2 was initially not confident in the system and scored 4 for this task, but developed his trust in the system later after asking several followup questions (Section \ref{R_questions}) to verify the changes, stating \QUOTE{I was uncertain whether the system was reliable so I gave you a neutral rating, but it seemed correct after I tested its reliability, so I could have gave a higher rating later.}
Lastly, though participants were able to increase the brightness of the dog (\(\mu\)=5.3, \(\sigma\)=1.1), {\name} occasionally described the white dog as a `cream-colored' or `light-colored' dog in \textsc{Updated Object Descriptions}, even though \textsc{Summary of Visual Changes} and \textsc{AI Judgement} both confirmed the changes of brightness. 
For instance, P3 specified \PROMPT{Make the dog brighter color} and scored five for his confidence after obtaining the updated object descriptions \QUOTE{Fluffy cream-colored dog sitting}, which made him confused: \QUOTE{I said to make it brighter color, which is maybe a little bit different, it's no longer white but cream color. I am not sure if cream is brighter than white.}

In session 3, most participants were able to perform and confirm the completion of the tasks, while a few (P2, P3, P6) encountered issues and had to skip certain tasks.
For example, P2 was unable to change the color of the bow tie to blue using his prompts, such as \PROMPT{change the color of the cat to white cat with blue bow tie} or \PROMPT{change the color of the cat's bow tie from red to blue}. It was because {\name} misclassified the object of interest as the cat rather than the bow tie and thus changed the color of the cat instead.
In another case, P3 encountered discrepant verification feedback or hallucinations after changing the bow tie to blue. For instance, in his two attempts, the bow tie became blue, but with object descriptions changed from \QUOTE{red bow tie with a bell...}, to \QUOTE{blue bow tie with a button or ornament...}, and \QUOTE{blue bow tie with polka dots...} 

In session 4, participants developed a number of editing tasks (\textit{e.g., }increasing the brightness of the sink or towel, blurring out the plastic jars) and were able to complete most of them, with a few failure cases encountered (P4, P8, P10).
P4 specified her prompt \PROMPT{Remove the pill bottle from this image}, in which the image had multiple bottles, but none of them was recognized as a `pill bottle' in the object descriptions, which made {\name} fail to ground on a specific object. 
In another example, after P8 increased the brightness of the towel, {\name} hallucinated \QUOTE{In the edited image, an object resembling a penguin has been added...}
P10 encountered \textsc{AI Judgment} indicating \QUOTE{the visual difference is indiscernible} while the \textsc{Updated Object Descriptions} for the tower changed from `dark' to `gray', which confused him.

\begin{figure*}[t]
\begin{center}
\includegraphics[width=\linewidth]{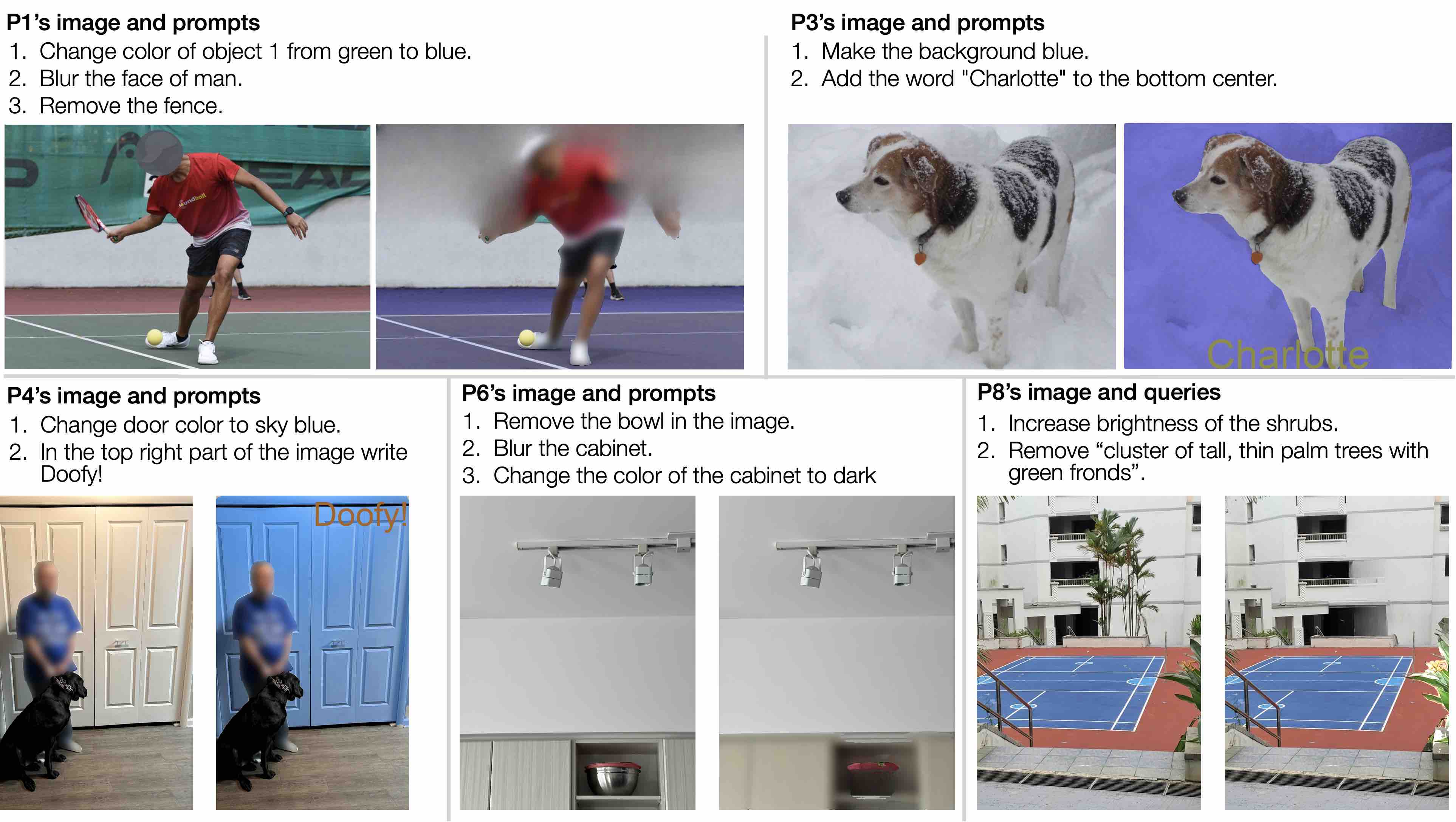}
\vspace{-1.5pc}
\caption{Participants' own images edited in study session 5, along with their prompts. \CHANGE{Faces are blurred for privacy.}
}
\label{fig:image_results}
\Description{P1’s image and queries
Change color of object 1 from green to blue.
Blur the face of man.
Remove the fence.
General description for input image:
A person playing tennis, wearing a red shirt, black shorts, a white cap, and a wristwatch is about to hit a low-flying tennis ball with a red and white racket, on a hard court with a green backdrop.
Object description for input image:
Object 1: "green tennis court with white boundary lines"
Object 2: "mesh fence with green covering"
Object 3: "gray wall"
Object 4: "man wearing a red t-shirt, black shorts, white sneakers, and a white cap playing tennis"
Object 5: "red and black tennis racket"
Object 6: "shadow on the court surface"
Object 7: "yellow tennis ball"
P3’s image and queries
Make the background blue.
Add the word "Charlotte" to the bottom center.
General description for input image: Dog with white and brown fur stands in deep snow, covered with snowflakes. It wears a collar with an orange tag and looks to the side. Snow continues to fall around it.
Object description for input image:
Object 1: "Snow-covered ground"
Object 2: "Dog with white fur and brown markings, wearing a collar with an orange tag, with snowflakes on its fur"
P4’s image and queries
Change door color to sky blue.
In the top right part of the image write Doofy!.
General description for input image: 
A man in a blue t-shirt and jeans stands indoors with his hand on the collar of a seated black dog. They are in front of white double doors, on a wooden floor. The man has facial hair and is wearing white socks.
Object description for input image:
Object 1: "set of white paneled doors with metallic handles"
Object 2: "wooden flooring with a greyish-brown color and plank design"
Object 3: "man with a beard wearing a blue graphic t-shirt and light-colored pants"
Object 4: "large black dog with a shiny coat and a bandana around its neck"
P6’s image and queries
Remove the bowl in the image.
Change the color of the cabinet to dark brown.
Blur the cabinet.
General description for input image: 
Two track lights attached to a ceiling rail above a kitchen cabinet with an open cubby displaying a metal bowl with a red lid. The wall and ceiling are painted in a light grey, and the cabinet has a wood grain texture.
Object descriptions for the input image: 
Object 1: "modern track lighting with two adjustable lights mounted on a horizontal rail"
Object 2: "light-colored wooden cabinet with textured vertical lines"
Object 3: "one of the adjustable lights on the track lighting fixture, aimed downwards"
Object 4: "silver stainless steel bowl with a visible red rim or lid"
P8’s image and queries
Increase brightness of the shrubs.
Remove “cluster of tall, thin palm trees with green fronds”.
General description for input image: An outdoor blue basketball court with white markings, surrounded by red pavement, tropical plants, and part of a residential building. A metal handrail and tiled steps lead down to the court area.
Object descriptions for the input image: 
Object 1: "outdoor sports court with blue and white markings for lines and boundaries"
Object 2: "white facade of a residential building with multiple balconies and decorative elements"
Object 3: "cluster of tall, thin palm trees with green fronds"
Object 4: "variety of smaller green plants and shrubs"
Object 5: "metal staircase railing painted black with a handrail and multiple steps leading downwards"
}
\end{center}
\end{figure*}

In session 5, five participants (P1, P3, P4, P6, P8) provided their own images (Figure \ref{fig:image_results}), and each performed two to three edits.
Most of the edits were successful, 
but there were a few instances where participants wanted to edit specific objects they ascertained but were not recognized by {\name} in sufficient granularity, \textit{e.g.,} the text on P1's T-shirt, P1's partially occluded watch, or the bowl lid by P6. 
Overall, participants were excited about the opportunities to edit their own images non-visually and came up with more desired edit actions and potential use cases in their daily lives.

\subsubsection{Promise of {\name} and need for finer controls and more edit actions}\label{R_finer_controls}
During the study, participants indicated a need for finer control over the edit actions.
First, participants expressed a desire to modify specific parts rather than the entire object.
Second, participants desired higher prompt bandwidth, such as editing multiple objects at once or using one detailed prompt to express desired and undesired edit outcomes (Section \ref{different_granularity}).
Participants also noted advanced functions, such as `moving object' (P10) or `applying and describing certain image filters' (P9).
Participants also sought higher granularity in adjustments, such as brightness, exact color codes, font family, and sizes.
This precise specification could help ensure that the system's output matches users' exact intent, as noted by P9 \QUOTE{AI is making its own decisions. It seems to complete the tasks but not my intent} (We will discuss more in Section \ref{generalizability}).

Also, participants generally appreciated that {\name} enabled them to edit images and had ideas for supporting everyday scenarios, such as making graphs or flyers for work (P5, P7), adding shapes or arrows to images to highlight content (P1, P2, P7, P8), or posting more images on their social media, as P4 said \QUOTE{If I get something like this in my hands, I will be taking a lot of pictures and I would totally be one of those people who take pictures of their food every time they make a meal.}
P7 also wanted to use {\name} to edit data graphs: \QUOTE{It's quite exciting to hear that such a system is being developed. As a data analyst, I program to create my graphs. It would help me to put a red box around the first bar to get people's attention, which I always find difficult and need someone to do it for me manually.} 
P9 further expressed his eagerness as a photographer and commended that using natural language could benefit more people: \QUOTE{I am trying to get back into taking photos because there are no accessible ways to do it. It could be helpful to describe the filter I want, and then it generates that filter and description. It could also be useful for even a layperson who doesn't have technical knowledge of editing software, where you can just use natural language to describe what you want and for the image to be returned in that way.}

\subsection{RQ2: How do BLV people prompt {\name}? }
\textit{Participants found it intuitive to use natural language to specify prompts and ask followup questions, which reduced their learning efforts. 
They also developed their own prompt strategies throughout the study.}

\subsubsection{Specifying prompts with varying levels of granularity for better clarity.}\label{different_granularity}
We observed that participants specified information at varied levels of granularity in their prompts. Their strategies evolved changeably based on the success of edits or verification feedback, yet the goal remained the same --- \textit{to ensure the system understood their prompts.}

Participants used simple prompts, such as \PROMPT{make the dog brighter} used by P9 in session 2, who noted: \QUOTE{That's very basic and concise. The idea is not open to interpretation. In theory, I don't think there should be much confusion about what this instruction means.}
As the study went on, he added more specificity to his prompts, evolving to \PROMPT{make the towel brighter color,} and eventually to \PROMPT{increase the brightness of the cat to make it more visible, without changing the color of the cat.} 
This demonstrated a progression towards more detailed instructions to better convey their intent and ensure the desired outcome.
However, several participants (P6, P8, P9, P10) developed more detailed information about their desired outcomes and specific exclusions in their edits initially. 
For instance, consider P6's prompt, who had image generation prompting experiences, in session 3: \PROMPT{Take the object of the cat and increase the brightness of the object so that there is greater contrast between the cat and the rest of the photo.}
However, facing failed edits or discrepancies in the verification feedback, P6 attributed those discrepancies to his complex prompts, and decided to simplify them to reduce potential confusion. 
For instance, when he edited his own image and aimed to remove a lid of the bowl, he changed his prompt from \QUOTE{Remove the lid in the image so that there is no presence of a bowl or any related accessories,} to \QUOTE{Remove the lid of the bowl,} and finally to a more succinct \QUOTE{Remove the lid in the image.}

We also observed that some participants conveyed high-level goals in their prompts to let the system interpret their intents and perform proper actions.
For example, in session 3, participants were asked to obscure the person in the image to maintain privacy and make a white cat noticeable. 
P7 articulated her prompt: \PROMPT{make the cream-colored cat brighter and stand out more in the image} and explained her rationale: \QUOTE{I added the second sentence, as my whole intention was to make it more pronounced.}
Additionally, some participants naturally crafted high-level prompts to express their overarching objectives for the scenario,  such as 
\PROMPT{I don't want to see the lady} (P1), 
\PROMPT{Increase the focus of the white cat} (P1), 
\PROMPT{Hide the bowl} (P3).
All these high-level prompts were linked correctly to the edit actions and objects.

For referencing the object of interest, participants generally chose to refer to objects by their names as given in the descriptions, such as `cat' or `woman.'
However, when errors occurred, they resorted to using detailed descriptors to ensure the system correctly identified the target object, for instance,
\PROMPT{Change the color of the white paneled wall to blue} (P6), or \PROMPT{Blur out the man in the framed photo within the image} (P9). 
Another option, in addition to object names, is to refer to indexes of objects according to the initial object descriptions. 
Some found referring to indexes of objects helpful in complex scenarios, such as in session 4, where P2 prompted \PROMPT{Blur out object 7} and mentioned \QUOTE{I had to admit the so many objects, makes it hard to digest. It's easier to refer to the number and use the natural language to describe what I want to blur out.}
However, some (P2, P9, P10) also expressed concerns about confusing the system, given that the object description may contain multiple objects as descriptors, which the system could incorrectly reference.

\subsubsection{Users' mental model affecting specified prompts.}\label{R_factors_prompt}
We observed that participants' mental model of how the {\name} system works influenced the way they formulated their prompts, varying in tone, use of symbols, and specific word choices.

Some participants anthropomorphized {\name}, treating it as a human-like agent. P3 perceived he was collaborating with human agents, stating, \QUOTE{two people were working on the image as if one was working on the image and another commentator commenting on the output}, and \QUOTE{It's not obedient, it's making its own decision.}
P4, similarly, adopted polite tones in their prompts as if she was interacting with someone, exemplified by the use of phrases such as ``Please...'' and ``Can you...'', and noted, \QUOTE{I wrote it the way I would speak to a friend editing this picture, like a conversational tone.}
Participants with extensive experience in visual arts and AI technology displayed a higher level of confidence and precision in their language.
P1, who used AI technologies in his daily life, showed this trust in {\name}'s inferential capabilities and prompted \PROMPT{Change object 1 from green to blue}. He assumed the AI would have general knowledge of tennis courts and maintain the necessary details, such as the white lines on the tennis court: \QUOTE{I assumed that AI would probably have taken some inference from other sources of what a blue tennis court is and generalized tennis courts to be blue color.} 

However, uncertainties about {\name}'s ability to understand nuanced linguistic cues led to concerns over how to format prompts effectively.  
P3 wondered if quotation marks were necessary to emphasize specific words within his instructions: \QUOTE{I thought I would have had to put the word puppy in quotation marks to indicate the beginning and end, like a tag.} 
Similarly, P6 was concerned that quotation marks included in his prompt might mistakenly appear in the final image and asked follow-up questions to confirm.
To distinguish between prompts that were questions and edit instructions, P5, who described himself as an intermediate programmer and tech-savvy, experimented and speculated about the system's method of recognizing prompts, \QUOTE{I was trying to figure out how you were determining whether it's a question or not, like if you're looking for a question mark, if you're doing a regular expression, or if you're having AI determine.}

\subsubsection{Objectives and strategies of prompting follow-up questions}\label{R_questions}
Participants often used follow-up questions to verify the results of their edits. 
For instance, after attempting to remove a bowl in session 2, participants asked \PROMPT{How many bowls are in the picture?} (P1), or \PROMPT{Is there any ceramic material in the picture?} (P9) to confirm that the removal was completed and no remnants were left behind.
This verification step was crucial to ensure the edits met their expectation when discrepancies arose in the outputs, as P9 remarked \QUOTE{I used summary and AI Judgement to compare to make sure that what I wanted was done. I was more likely to ask follow-up questions when the 2 of them were inconsistent.}
Also, participants were concerned with the aesthetic quality of the images post-edit. 
They inquired about the overall appearance and integrity of the edits by asking questions such as \PROMPT{Does the image look doctored?} (P5), \PROMPT{Do you see any imperfections in the photo?} (P5), or \PROMPT{Describe the brightness of the dog in contrast to the rest of the image.} (P9).

Asking for spatial information was also prevalent, especially when participants planned to add text to an image or assess the text they added.
They sought to understand the layout and space availability, such as 
\PROMPT{Are there areas in the image where it's empty, or less cluttered?} (P10), or
\PROMPT{What is the size of the word ``puppy'' compared to the whole of the picture?} (P2).

Lastly, participants prompted {\name} of their existing knowledge to gain confidence and gauge the system's reliability. 
For instance, after removing a bowl, P2 systematically prompted {\name} with questions he knew the answers to, such as \PROMPT{Is there a bowl in the picture?} (answer: \PROMPT{No}), \PROMPT{Is there a dog in the picture?} (answer: \PROMPT{Yes}), \PROMPT{Is there a man in the picture?} (answer: \PROMPT{No}). 
He developed this confidence in {\name} by confirming known facts, and believed the bowl was removed \QUOTE{I have doubts about whether the bowl was removed, and then to further try and get more confidence as to whether the AI can answer the question correctly. The first question is to see whether the bowl is removed. The other two questions are just to check whether the AI is reliable.}

\subsection{RQ3: How do BLV people perceive {\name}'s verification feedback?}\label{RQ_perceive_output}
\textit{Participants found all four types of verification feedback useful depending on the context and information they wanted.
They developed different strategies and interpretations for consuming verification feedback.}

\subsubsection{Perception on each verification feedback.}\label{R_perceive_output}
We reported participants' perceptions of each verification feedback.

\textbf{Summary of Visual Changes:} 
Participants perceived \textsc{Summary of Visual Changes} as `clear' (P7) and `direct' (P1) into the actions that have been performed.
P4 commented \QUOTE{Summary is really good at comparing the original and the latest image for the changes. So that's very helpful in giving me an idea of where I was, and this is where I am.}
However, participants felt that the summary alone might be too superficial and lacked detailed reasoning to fully assess the success of the edits. Therefore, participants combined it with the other feedback.

\textbf{AI Judgement:} 
Based on the above reason, participants tended to combine and compare \textsc{Summary of Visual Changes} and \textsc{AI Judgement}, which they perceived the former as the facts, while the latter as \QUOTE{internal thinking and reasoning of the system} (P5). P1 indicated \QUOTE{AI judgment gives a more detailed description into what you actually assess the thing, even more descriptively which part has been changed.} 
However, some (P5, P7) regarded the information of \textsc{AI Judgement} as verbose and overlapped from other outputs, as P5 suggested \QUOTE{There might be a point where they become redundant. But at the moment, I appreciate having them all.}

\textbf{Updated General Descriptions:} 
Most participants found \textsc{Updated General Descriptions} useful for gaining a new perspective on how `sighted individuals' may perceive the image, as highlighted by the terms \QUOTE{new angle} (P2) and \QUOTE{new mental image} (P1).  
Also, as P9 noted, \QUOTE{general descriptions determine how much focus that object has when it comes to someone seeing the picture. So it allows the user to construct a better visual idea of what the picture is like.}
Most participants also mentioned that general descriptions offer more global and comprehensive information, enabling them to picture the image content, as P3 described \QUOTE{It describes everything in relation to the different objects in relation to each other. It tells me what the image looks like.}
However, general descriptions were generated based solely on the updated image, and the system could generate varying descriptions for the unchanged objects, which confused the user (Section \ref{R_consistency_discrepancy}).

\textbf{Updated Object Descriptions:} 
These were perceived as useful for noting detailed edits and serving as a reference for prompts. P1 highlighted this benefit: \QUOTE{It actually helps me to precisely call out the object so that I can speak this language. I can know how it actually is. I can actually make the amendment.} 
However, object descriptions in session 4 had rich information on granular objects in their descriptors, as well as speculative information (\textit{e.g.,}`possibly for...', `likely filled with...', `potentially a ...' ), which made them confusing to absorb, as noted by P9: \QUOTE{The quantity of objects is confusing. I wasn't sure which object I was manipulating, because it seemed like there were more objects than there actually were.}

\subsubsection{Consistency vs. discrepancy among the verification feedback.}\label{R_consistency_discrepancy}
Generally, participants were confident in an edit if all the feedback was consistent. 
P7 stated \QUOTE{
Summary and AI judgement confirm object 5 has been removed, so it just put object 5 a colon in object description. I also don't see the bowl in the general description anymore. I think it's correct as I have multiple avenues of confirmation.}
However, we observed occasional discrepancies in the feedback, such as different types of feedback describing the same object in different ways, prompting participants to seek clarification by asking follow-up questions.
For instance, P7, who frequently evaluated and compared \textsc{Summary of Visual Changes} and \textsc{AI Judgement}, thought aloud when tackling a discrepancy: \QUOTE{The summary and AI judgment don't really correspond, because summary says the person has been removed, and AI judgment says the woman can no longer be clearly identified. So I think that there's some discrepancy here. I guess I should check. [prompting `Is there a blurred image of human the in background' (answer `yes')] So it seems like the summary was wrong, and the judgment was correct.} 
Discrepancies were also noted between the current and previous verification feedback.
For instance, after P3 changed the color of a cat's bow tie from red to blue, the system feedback added an unexpected detail: polka dots. P3 noted, \QUOTE{It was like making its own decisions to add polka dots. I don't know if these were there. Just wasn't describing it before.}
P9 also mentioned, \QUOTE{One time it described as teal, one time described as greenish gray. I believe it was describing the same container. So, making sure to use the same kind of descriptors would be quite helpful.}

\subsubsection{Difficulty picturing editing effects leads to challenges judging edit outputs.}\label{R_unknown_output}
During the study, participants expressed difficulties in picturing certain edited visual effects due to their lack of visual experience, 
echoed with findings of prior work~\cite{lotusChi24} around the challenge of envisioning complex obfuscation effects (\textit{e.g.,} inpainting an area with surrounding background) by BLV individuals.
For instance, P2 was uncertain about the effect of increasing brightness: \QUOTE{I have to admit I'm not sure what is supposed to happen when I make something brighter.}
Similarly, P6 was confused by the effects of brightening a white dog during session 2, which resulted in the color description changing from white to cream in the new verification feedback. 
P6 noted \QUOTE{I think there wasn't really anything in the description to explain why it chose that particular color or whether increasing the brightness naturally changes it to this color.}
Some participants (P3, P5) also struggled to grasp the intended result of blurring effects on images, as P5 commented \QUOTE{as a blind person. I'm not entirely sure what Blurred is supposed to do.}

\subsubsection{\CHANGE{Excessive information and tone influence the perception and confidence of the results.}}
Within the verification feedback, {\name} sometimes generated details excessive to what users required or presented information using a less confident tone, leading to confusion among participants.
For instance, P2 encountered an object description of a man stated as \textit{``Framed portrait of a person with nondescript features due to blurring for privacy.''} P2 found this confusing, as he did not instruct the system to do it for privacy purposes: \QUOTE{It's a bit strange. That was totally not in my instruction at all. It provides extra information. Overly explain things.}
Moreover, some participants were also influenced by the negative tones in the verification feedback, which they thought decreased their confidence in the results.
For example, P1 was unsure about the success of an edit due to the uncertain tone of the feedback: \QUOTE{It said the edit appears to be successful, so is this successful or not?}
In contrast, a positive tone in the outputs could enhance confidence, as noted by P3 \QUOTE{The system seems to be pretty confident, saying the edit is successful. Just based purely on that confident, sounding language. I am willing to give it some credit.}

\subsection{RQ4: How do BLV people perceive the final edited images?}
\textit{Participants were generally willing to use the final edited image depending on the scenario but expressed the need for further validation from other sources.}

\subsubsection{Determining whether to post on public depending on context.}
Participants generally expressed confidence that the system had successfully completed the tasks as instructed, though they were uncertain if it had fully met their intentions.
P9, with visual experiences and higher expectations for visual aesthetics, commented \QUOTE{It was successful in actually doing what I asked it to, but maybe not the intent. I just wanted to know the nature of the text that was added. It's good that it told me it's San Serif, but it's not very useful in a practical way, like the aesthetic quality of the text.}
Consequently, due to concerns about the unseen visual outputs, most participants preferred to seek verification from sighted assistance before posting, as suggested by P1 \QUOTE{I will probably get somebody just to help me to visually check and confirm before I actually post it. I think there's a certain level of trust in it. But it's not 100\%.}

However, some participants noted that they might consider publishing the images in urgent situations or if sighted assistance was unavailable since they perceived the system satisfied the basic requirements. 
For instance, P7 mentioned the scenario of looking for the cat, where urgency overrides others: \QUOTE{If I feel that it's very urgent, and there's no one else to check over it, then I will go ahead and post because I know that the picture shows my cat, and it shows my number.}
Also, P9, though desired the visual check as mentioned, acknowledged the primary goal of the edit over nuanced visual quality, 
\QUOTE{The quality of the edit is secondary. Because what you really need is for people to pay attention to the photo of the cat, to make sure that it's recognizable and distinct, and also to have the call to this number. And based on what the AI judgment is telling me, it seems like it performed the task that I asked it to.}

\subsubsection{Need for multiple modalities or sources to verify output images}\label{R_multiple_modalities}
In addition to seeking sighted assistance, participants employed various strategies to verify the content of images. 
Many participants were tech-savvy at utilizing AI technologies in their daily routines, including BeMyAI \cite{bemyai}, SeeingAI \cite{seeingai}, Midjourney \cite{midjourney}, and ChatGPT \cite{chatgpt}. 
Therefore, a common method among them was to use different image captioning models or applications to cross-verify the image content, echoing findings from Herskovitz et al. \cite{diyat} on how BLV people combined and utilized assistive technologies.
P6, who usually used an embosser or Braille display to check the content and spatial information of images he created with Midjourney: \QUOTE{I have an embosser that I can send images to. So I would probably send images to it before posting online. I could recognize something supposedly pretty simplistic, like a ball or a dog, and check the spatial relationship.}
For the tasks for adding text, participants also mentioned verifying the text further using optical character recognition (OCR) built into the screen reader or other third-party apps.
P6 further pointed out using sonification to make those changes of color or brightness accessible.

\subsubsection{Faith and experiences with AI technologies affect the decision of using final edited images}\label{R_ai_experiences}
Half participants (P2, P4, P5, P6, P9) expressed a lack of confidence not only in {\name} but also in other AI technologies, stemming from the imperfections they had observed in their experiences. 
P9, for example, shared his skepticism based on previous experiences with image editing software, particularly with object removal: \QUOTE{I am less inclined to do it because I do not fully trust that it will do it reliably. But for other functions that make a mistake, it wouldn't be a terrible one.} 
Similarly, P4 discussed her overall trust in AI technologies, emphasizing the inherent uncertainties faced by BLV users: \QUOTE{
Being totally blind, you could be told the sky is green and the grass is purple. There's no way for me to technically say that's not true. There's an element of you having to put trust into a system and hope it's doing what it's supposed to be doing. It's not really a concern but a lifestyle. It's part of being blind. You have to put a lot of faith in the technology you use and hope it's doing what it's supposed to.}
In contrast, we found that participants familiar with AI technologies tended to trust the verification feedback more, and recommended simplifying the information {\name} provided. 
P7, a data analyst who regularly programmed and created visualizations for work and used off-the-shelf AI-enabled image captioning apps, suggested: \QUOTE{I use all these AI every day for my work. We are used to AI, and it worked as it should in general. I think the system [{\name}] should work this way. Then I think there's no need for so many levels of checks at verification. So most people will actually be fine, with just a short description of the change, and say this to be successful.}

\section{Discussion and Future Work}
In this section, we discuss our limitations, lessons learned, and future work to support non-visual image editing.

\subsection{Enhancing Verification Feedback Loops}
While the verification feedback was generally perceived as useful in providing different aspects of edit information, there are many opportunities to make them more informative and personalized to BLV users.

First, although {\name}'s natural language verification loops aim to support BLV image editors' diverse preferences for visual feedback~\cite{stangl21assets,lotusChi24} and did prove to bring more flexibility (Section \ref{R_questions}), we noted a need for further customization, such as combining or omitting certain elements for greater conciseness (Section \ref{R_perceive_output}), elaborating specific visual effects (Section \ref{R_unknown_output}), incorporating detailed spatial descriptions (Section \ref{R_questions}).
It is also important to consider BLV users' prior visual experiences and experiences with AI technologies (Section \ref{R_ai_experiences}), which could provide them with proper levels of feedback \textit{e.g.,} providing professional terms for describing visuals accurately (as for P9 who is a photographer), or providing succinct and direct feedback if users are familiar with and trust AI technologies (as for P7 who used GenAI for daily work).

Second, participants were unclear about the linguistic and editing capabilities of {\name}. 
This confusion led to repetitive prompt refinement through trial and error. 
Drawing on prior research that emphasizes the importance of explaining an automated system's decisions \cite{nourani2019effects,wu2017, zhao2017, macleod2017, amershi2019}, 
the system could clarify its capabilities of interpreting natural language prompts into specific items (\textit{e.g.,} edit functions, objects), thereby aiding in establishing a user’s initial mental model of the system.
The system can also improve its effectiveness by building a model of the user, analyzing user prompts, and initiating follow-up questions to elicit user needs and goals. 
\CHANGE{After understanding the user's needs, the system can assign corresponding agents, who may have different domain knowledge, such as interior design, graph design, etc. These agents could provide more efficient and useful feedback tailored to the user's needs and goals.}
This mutual learning and adaptive approach could allow the system to personalize the visual verification feedback to the user progressively.

Occasional hallucinations and inconsistent descriptors across verification feedback are other significant issues. 
Potential solutions include incorporating historically-used wordings into prompts for generating consistent feedback, and strategically constraining and guiding the lexical choices of LLMs \cite{zeng2023conzic,konen2024style} (\textit{e.g.,} tuning the temperature parameter of GPT). 
It can also be beneficial to include different VLMs and LLMs to cross-verify the edit and provide verification feedback from different perspectives, as how BLV people suggested and used in their current workflow of verifying visual information (Section \ref{R_multiple_modalities} \& \cite{diyat,genassist}). 
\CHANGE{Sighted people can also be involved in the image editing process if needed, as prior work has shown different techniques to support collaborative visual \cite{diffscriber,zhang2023imageally} or text editing \cite{collaba11y,collab}.}
Furthermore, techniques in different modalities have been proposed and utilized to verify visual content, (\textit{e.g.,} tactile images \cite{a11yboard, li2019, lee2022imageexplorer, fan2022}, non-speech sounds~\cite{siu2022, a11yboard}). 
Future work could explore ways to integrate these methods to offer a more consistent and reliable experience for BLV users. 

\subsection{Supporting  Richer Edit Actions}\label{generalizability}
In this paper, we demonstrated the utility of natural language verification loops through {\name} that supports five edit actions. 
\CHANGE{Our participants expressed the need to have finer controls and more edit actions (Section \ref{RQ1_results})}, which can be achieved in the future due to {\name}'s flexible structure. 
For instance, with the power of LLMs, {\name} can understand more edit actions and allow users to specify a particular value to a certain visual effect (\textit{e.g.,} changing color with color codes) by few-shot prompting (\textit{e.g.,} giving a series of examples as the context of what the model should perform) \cite{brown2020language, yuan2022wordcraft}, which could further increase the bandwidth of input prompt and users can perform multiple edits at a time.
We can also utilize the models of different granular object segmentation to support finer controls on parts of objects~\cite{seem, kirillov2023segment, li2023semantic}, similar to techniques that enable hierarchical image explorations~\cite{lee2022imageexplorer, imageexplorerw4a, seeingai}. 
\CHANGE{Other models could enable additional actions, such as recognizing shapes to enable diagram editing, or detecting depth to enable occlusion-aware editing.
In contrast to finer edits, another potential integration is to include models that enable global content editing, such as InstructPix2Pix~\cite{brooks2023instructpix2pix} to adjust the overall style of the image or Emu Edit~\cite{sheynin2023emu} to generate new content consistent with the overall image.
Unlike traditional image editing, such generative edits often consider the overall image content and optimize the visual aesthetic, which could create unintended results. 
}
Therefore, future work could explore how to describe such global or partial style changes in verification feedback. 
This should cover not only the content of the images but also any perceived imperfections or artifacts in the edited images (\textit{e.g., }text cutoffs, visual artifacts).

\subsection{Limitations of the Study}
In our user study, we tried to recruit BLV individuals with diverse backgrounds. Some were familiar with AI technologies, some had extensive visual experiences, and some had image editing experiences. 
However, our study insights may not represent sufficiently broad perspectives in the BLV community, who have varied experiences and use of technologies.
Future studies could explore this broader
set of experiences and perspectives through field studies or deployment. 
Second, we selected specific images across different sessions to showcase various levels of image complexity and practical contexts. However, these chosen images may not accurately represent the types of photos that BLV individuals capture in their daily lives. 
Although participants in our study session 5 provided and edited their own images, the small sample may not represent the range of characteristics taken by blind people.
Future work could further explore how {\name} would perform with images BLV people personally capture and how they would manage editing and iterating these photos on their own with {\name}. 

\section{Conclusion}
We have presented {\name}, a prototype system demonstrating the concept of natural language verification loops in the context of image editing.
{\name} supports five specific edit actions and allows BLV users to input their editing instructions in natural language.
Building on insights from prior work, we developed four types of verification feedback for BLV users to perceive changes made to images. 
These include a summary of visual changes, AI judgments, and general and object descriptions.
Through a user study with ten BLV people, we explored how {\name} supported non-visual image editing, how participants prompted {\name}, and how they perceived verification feedback and the final edited images. 
Finally, we discussed ways to enhance the verification feedback provided by {\name}, explored possibilities for expanding the range of edit actions available, and outlined implications for future systems that utilize natural language verification loops. 

\begin{acks}
We thank our anonymous reviewers and all the participants in our study for their suggestions, as well as Andi Xu for helping facilitate our user studies.
\end{acks}

\bibliographystyle{ACM-Reference-Format}
\bibliography{imageio}

\appendix

\begin{figure*}[t]
\vspace{1pc}
\begin{center}
\includegraphics[width=0.95\linewidth]{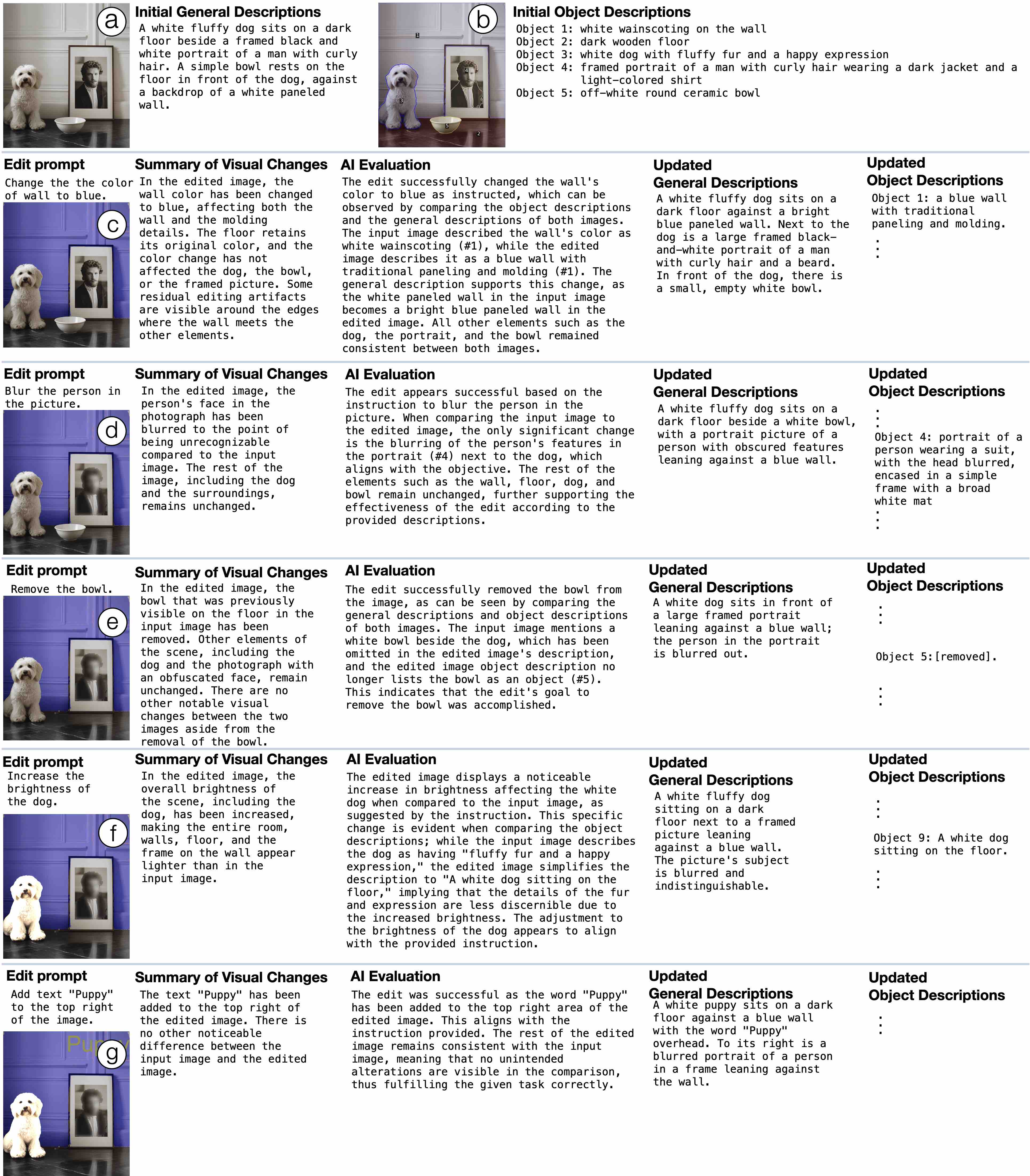}
\vspace{-0.5pc}
\caption{Detailed edit prompts and corresponding verification feedback for the Walkthrough and Session 2 in our study.  
}
\vspace{1pc}
\label{fig:dog_scenario}
\Description{This figure demonstrate the verification output corresponding to certain edit query in session 2.
Here is the initial general descriptions of the image: A white fluffy dog sits on a dark floor beside a framed black and white portrait of a man with curly hair. A simple bowl rests on the floor in front of the dog, against a backdrop of a white paneled wall.
Here is the initial object descriptions of the image: 
Object 1: white wainscoting on the wall
Object 2: dark wooden floor
Object 3: white dog with fluffy fur and a happy expression
Object 4: framed portrait of a man with curly hair wearing a dark jacket and a
 light-colored shirt
Object 5: off-white round ceramic bowl
Here is the verification feedback of the edit query 1: Change the the color of wall to blue.
Summary of visual changes: In the edited image, the wall color has been changed to blue, affecting both the wall and the molding details. The floor retains its original color, and the color change has not affected the dog, the bowl, or the framed picture. Some residual editing artifacts are visible around the edges where the wall meets the other elements.
AI judgment: The edit successfully changed the wall's color to blue as instructed, which can be observed by comparing the object descriptions and the general descriptions of both images. The input image described the wall's color as white wainscoting (#1), while the edited image describes it as a blue wall with traditional paneling and molding (#1). The general description supports this change, as the white paneled wall in the input image becomes a bright blue paneled wall in the edited image. All other elements such as the dog, the portrait, and the bowl remained consistent between both images.
Updated general description: A white fluffy dog sits on a dark floor against a bright blue paneled wall. Next to the dog is a large framed black-and-white portrait of a man with curly hair and a beard. In front of the dog, there is a small, empty white bowl.
Updated object description: Object 1: a blue wall with traditional paneling and molding.
Here is the verification feedback of the edit query 2: Blur the person in the picture.
Summary of visual changes: In the edited image, the person's face in the photograph has been blurred to the point of being unrecognizable compared to the input image. The rest of the image, including the dog and the surroundings, remains unchanged.
AI judgment: The edit appears successful based on the instruction to blur the person in the picture. When comparing the input image to the edited image, the only significant change is the blurring of the person's features in the portrait (#4) next to the dog, which aligns with the objective. The rest of the elements such as the wall, floor, dog, and bowl remain unchanged, further supporting the effectiveness of the edit according to the provided descriptions.
Updated general description: A white fluffy dog sits on a dark floor beside a white bowl, with a portrait picture of a person with obscured features leaning against a blue wall.
Updated object description: Object 4: portrait of a person wearing a suit, with the head blurred, encased in a simple frame with a broad white mat.
Here is the verification feedback of the edit query 3: Remove the bowl.
Summary of visual changes: In the edited image, the bowl that was previously visible on the floor in the input image has been removed. Other elements of the scene, including the dog and the photograph with an obfuscated face, remain unchanged. There are no other notable visual changes between the two images aside from the removal of the bowl.
AI judgment: The edit successfully removed the bowl from the image, as can be seen by comparing the general descriptions and object descriptions of both images. The input image mentions a white bowl beside the dog, which has been omitted in the edited image's description, and the edited image object description no longer lists the bowl as an object (#5). This indicates that the edit's goal to remove the bowl was accomplished.
Updated general description: A white dog sits in front of a large framed portrait leaning against a blue wall; the person in the portrait is blurred out.
Updated object description: Object 5:[removed].
Here is the verification feedback of the edit query 4: Increase the brightness of the dog.
Summary of visual changes: In the edited image, the overall brightness of the scene, including the dog, has been increased, making the entire room, walls, floor, and the frame on the wall appear lighter than in the input image.
AI judgment: The edited image displays a noticeable increase in brightness affecting the white dog when compared to the input image, as suggested by the instruction. This specific change is evident when comparing the object descriptions; while the input image describes the dog as having "fluffy fur and a happy expression," the edited image simplifies the description to "A white dog sitting on the floor," implying that the details of the fur and expression are less discernible due to the increased brightness. The adjustment to the brightness of the dog appears to align with the provided instruction.
Updated general description: A white fluffy dog sitting on a dark floor next to a framed picture leaning against a blue wall. The picture's subject is blurred and indistinguishable.
Updated object description: Object 9: A white dog sitting on the floor.
Here is the verification feedback of the edit query 5: Add text "Puppy" to the top right of the image.
Summary of visual changes: The text "Puppy" has been added to the top right of the edited image. There is no other noticeable difference between the input image and the edited image.
AI judgment: The edit was successful as the word "Puppy" has been added to the top right area of the edited image. This aligns with the instruction provided. The rest of the edited image remains consistent with the input image, meaning that no unintended alterations are visible in the comparison, thus fulfilling the given task correctly.
Updated general description: A white puppy sits on a dark floor against a blue wall with the word "Puppy" overhead. To its right is a blurred portrait of a person in a frame leaning against the wall.
Updated object description: None
}
\end{center}
\end{figure*}

\begin{figure*}[t]
\vspace{1pc}
\begin{center}
\includegraphics[width=0.95\linewidth]{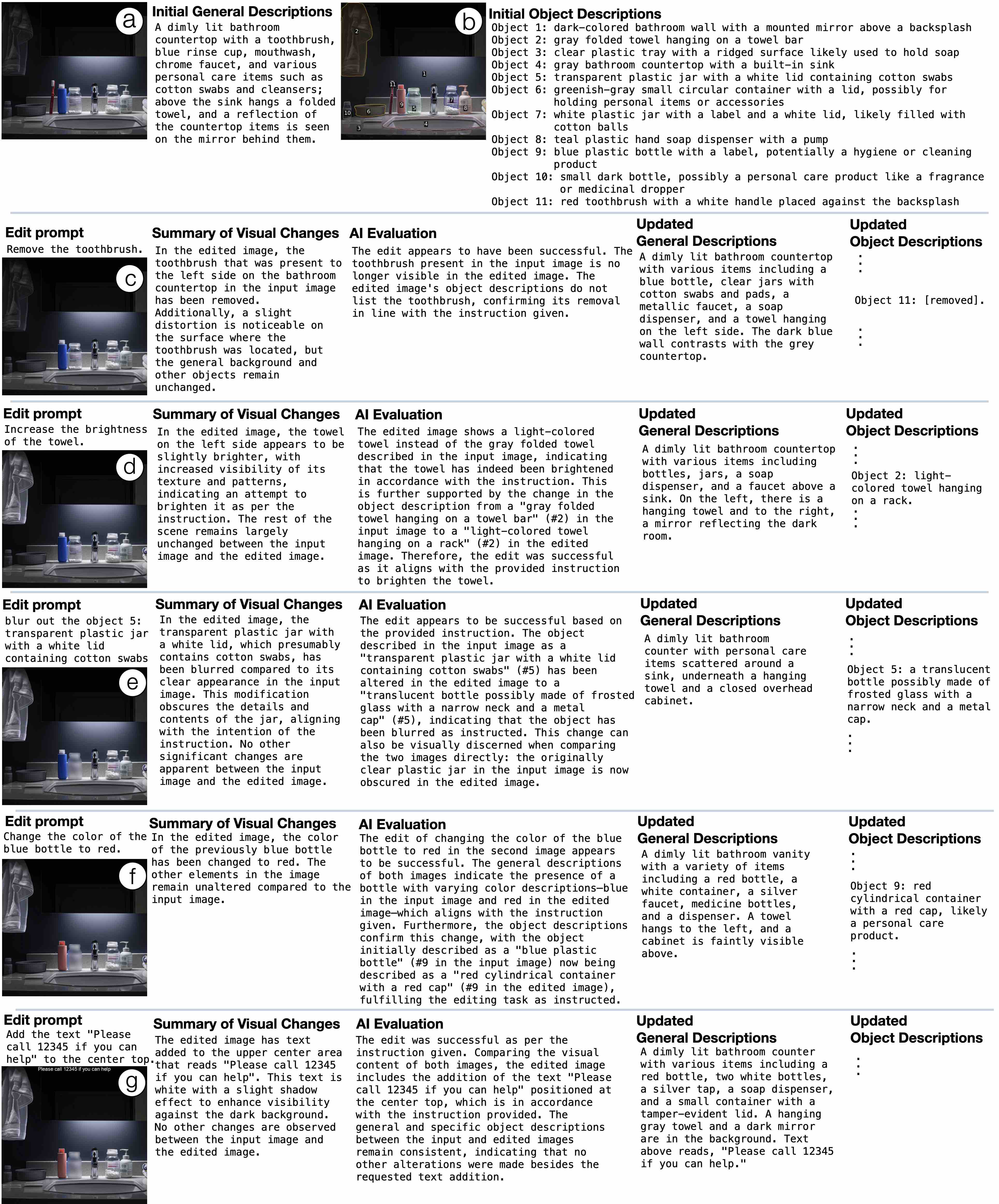}
\vspace{-0.5pc}
\caption{Detailed edit prompts and corresponding verification feedback for the Walkthrough and Session 4 in our study.  
}
\vspace{1pc}
\label{fig:private_scenario}
\Description{This figure demonstrate the verification output corresponding to certain edit query in session 4.
Here is the initial general descriptions of the image: A dimly lit bathroom countertop with a toothbrush, blue rinse cup, mouthwash, chrome faucet, and various personal care items such as cotton swabs and cleansers; above the sink hangs a folded towel, and a reflection of the countertop items is seen on the mirror behind them.
Here is the initial object descriptions of the image: 
Object 1: dark-colored bathroom wall with a mounted mirror above a backsplash
Object 2: gray folded towel hanging on a towel bar
Object 3: clear plastic tray with a ridged surface likely used to hold soap
Object 4: gray bathroom countertop with a built-in sink
Object 5: transparent plastic jar with a white lid containing cotton swabs
Object 6: greenish-gray small circular container with a lid, possibly for holding personal items or accessories
Object 7: white plastic jar with a label and a white lid, likely filled with cotton balls
Object 8: teal plastic hand soap dispenser with a pump
Object 9: blue plastic bottle with a label, potentially a hygiene or cleaning product
Object 10: small dark bottle, possibly a personal care product like a fragrance or medicinal dropper
Object 11: red toothbrush with a white handle placed against the backsplash
Here is the verification feedback of the edit query 1: Remove the toothbrush.
Summary of visual changes: In the edited image, the toothbrush that was present to the left side on the bathroom countertop in the input image has been removed. Additionally, a slight distortion is noticeable on the surface where the toothbrush was located, but the general background and other objects remain unchanged.
AI judgment: The edit appears to have been successful. The toothbrush present in the input image is no longer visible in the edited image. The edited image's object descriptions do not list the toothbrush, confirming its removal in line with the instruction given.
Updated general description: A dimly lit bathroom countertop with various items including a blue bottle, clear jars with cotton swabs and pads, a metallic faucet, a soap dispenser, and a towel hanging on the left side. The dark blue wall contrasts with the grey countertop.
Updated object description: Object 11: [removed].
Here is the verification feedback of the edit query 2: Increase the brightness of the towel.
Summary of visual changes: In the edited image, the towel on the left side appears to be slightly brighter, with increased visibility of its texture and patterns, indicating an attempt to brighten it as per the instruction. The rest of the scene remains largely unchanged between the input image and the edited image.
AI judgment: The edited image shows a light-colored towel instead of the gray folded towel described in the input image, indicating that the towel has indeed been brightened in accordance with the instruction. This is further supported by the change in the object description from a "gray folded towel hanging on a towel bar" (#2) in the input image to a "light-colored towel hanging on a rack" (#2) in the edited image. Therefore, the edit was successful as it aligns with the provided instruction to brighten the towel.
Updated general description: A dimly lit bathroom countertop with various items including bottles, jars, a soap dispenser, and a faucet above a sink. On the left, there is a hanging towel and to the right, a mirror reflecting the dark room.
Updated object description: Object 2: light-colored towel hanging on a rack.
Here is the verification feedback of the edit query 3: blur out the object 5: transparent plastic jar with a white lid containing cotton swabs
Summary of visual changes: In the edited image, the transparent plastic jar with a white lid, which presumably contains cotton swabs, has been blurred compared to its clear appearance in the input image. This modification obscures the details and contents of the jar, aligning with the intention of the instruction. No other significant changes are apparent between the input image and the edited image.
AI judgment: The edit appears to be successful based on the provided instruction. The object described in the input image as a "transparent plastic jar with a white lid containing cotton swabs" (#5) has been altered in the edited image to a "translucent bottle possibly made of frosted glass with a narrow neck and a metal cap" (#5), indicating that the object has been blurred as instructed. This change can also be visually discerned when comparing the two images directly: the originally clear plastic jar in the input image is now obscured in the edited image.
Updated general description: A dimly lit bathroom counter with personal care items scattered around a sink, underneath a hanging towel and a closed overhead cabinet.
Updated object description: Object 5: a translucent bottle possibly made of frosted glass with a narrow neck and a metal cap.
Here is the verification feedback of the edit query 4: Change the color of the blue bottle to red.
Summary of visual changes: In the edited image, the color of the previously blue bottle has been changed to red. The other elements in the image remain unaltered compared to the input image.
AI judgment: The edit of changing the color of the blue bottle to red in the second image appears to be successful. The general descriptions of both images indicate the presence of a bottle with varying color descriptions—blue in the input image and red in the edited image—which aligns with the instruction given. Furthermore, the object descriptions confirm this change, with the object initially described as a "blue plastic bottle" (#9 in the input image) now being described as a "red cylindrical container with a red cap" (#9 in the edited image), fulfilling the editing task as instructed.
Updated general description: A dimly lit bathroom vanity with a variety of items including a red bottle, a white container, a silver faucet, medicine bottles, and a dispenser. A towel hangs to the left, and a cabinet is faintly visible above.
 Updated object description: Object 9: red cylindrical container with a red cap, likely a personal care product.
Here is the verification feedback of the edit query 5: Add the text "Please call 12345 if you can help" to the center top.
Summary of visual changes: The edited image has text added to the upper center area that reads "Please call 12345 if you can help". This text is white with a slight shadow effect to enhance visibility against the dark background. No other changes are observed between the input image and the edited image.
AI judgment: The edit was successful as per the instruction given. Comparing the visual content of both images, the edited image includes the addition of the text "Please call 12345 if you can help" positioned at the center top, which is in accordance with the instruction provided. The general and specific object descriptions between the input and edited images remain consistent, indicating that no other alterations were made besides the requested text addition.
Updated general description: A dimly lit bathroom counter with various items including a red bottle, two white bottles, a silver tap, a soap dispenser, and a small container with a tamper-evident lid. A hanging gray towel and a dark mirror are in the background. Text above reads, "Please call 12345 if you can help."
Updated object description: None
}
\end{center}
\end{figure*}

\end{document}